\documentclass[twocolumn,aps,showpacs,floats]{revtex4}

\usepackage{graphicx}
\usepackage{dcolumn}
\usepackage{bm}
\usepackage{epstopdf}

\newcommand{\bq}{\begin{equation}}
\newcommand{\eq}{\end{equation}}
\newcommand{\bqa}{\begin{eqnarray}}
\newcommand{\eqa}{\end{eqnarray}}
\newcommand{\nn}{\nonumber \\}

\def\be     {\begin{equation}}
\def\ee     {\end{equation}}
\def\bea        {\begin{eqnarray}}
\def\eea        {\end{eqnarray}}
\def\bnn    {\begin{eqnarray*}}
\def\enn    {\end{eqnarray*}}

\begin{document}

\title{Competition between Kondo and RKKY correlations in the presence of strong randomness}

\author{Minh-Tien Tran$^{1,2}$ and Ki-Seok Kim$^{1,3}$}
\affiliation{$^1$Asia Pacific Center for Theoretical Physics,
Pohang, Gyeongbuk 790-784, Republic of Korea \\
$^2$Institute of Physics, Vietnamese Academy of Science and
Technology, P.O.Box 429, 10000 Hanoi, Vietnam \\ $^3$Department of
Physics, POSTECH, Pohang, Gyeongbuk 790-784, Korea}

\begin{abstract}
We propose that competition between Kondo and magnetic
correlations results in a novel universality class for heavy
fermion quantum criticality in the presence of strong randomness.
Starting from an Anderson lattice model with disorder, we derive
an effective local field theory in the dynamical mean-field theory
(DMFT) approximation, where randomness is introduced into both
hybridization and Ruderman-Kittel-Kasuya-Yosida (RKKY)
interactions. Performing the saddle-point analysis in the U(1)
slave-boson representation, we reveal its phase diagram which
shows a quantum phase transition from a spin liquid state to a
local Fermi liquid phase. In contrast with the clean limit of the
Anderson lattice model, the effective hybridization given by holon
condensation turns out to vanish, resulting from the zero mean
value of the hybridization coupling constant. However, we show
that the holon density becomes finite when variance of
hybridization is sufficiently larger than that of the RKKY
coupling, giving rise to the Kondo effect. On the other hand, when
the variance of hybridization becomes smaller than that of the
RKKY coupling, the Kondo effect disappears, resulting in a fully
symmetric paramagnetic state, adiabatically connected with the
spin liquid state of the disordered Heisenberg model. We
investigate the quantum critical point beyond the mean-field
approximation. Introducing quantum corrections fully
self-consistently in the non-crossing approximation, we prove that
the local charge susceptibility has exactly the same critical
exponent as the local spin susceptibility, suggesting an enhanced
symmetry at the local quantum critical point. This leads us to
propose novel duality between the Kondo singlet phase and the
critical local moment state beyond the Landau-Ginzburg-Wilson
paradigm. The Landau-Ginzburg-Wilson forbidden duality serves the
mechanism of electron fractionalization in critical impurity
dynamics, where such fractionalized excitations are identified
with topological excitations.
\end{abstract}


\maketitle

\section{Introduction}

Interplay between interactions and disorders has been one of the
central issues in modern condensed matter physics
\cite{Interaction_Disorder_Book,RMP_Disorder_Interaction}. In the
weakly disordered metal the lowest-order interaction-correction
was shown to modify the density of states at the Fermi energy in
the diffusive regime \cite{AAL}, giving rise to non-Fermi liquid
physics particulary in low dimensions less than $d = 3$ while
further enhancement of electron correlations was predicted to
cause ferromagnetism \cite{Disorder_FM}. In an insulating phase
spin glass appears ubiquitously, where the average of the spin
moment vanishes in the long time scale, but local spin
correlations become finite, making the system away from
equilibrium \cite{SG_Review}.

An outstanding question is the role of disorder in the vicinity of
quantum phase transitions
\cite{Disorder_QCP_Review1,Disorder_QCP_Review2}, where effective
long-range interactions associated with critical fluctuations
appear to cause non-Fermi liquid physics
\cite{Disorder_QCP_Review2,QCP_Review}. Unfortunately, complexity
of this problem did not allow comprehensive understanding until
now. In the vicinity of the weakly disordered ferromagnetic
quantum critical point, an electrical transport-coefficient has
been studied, where the crossover temperature from the ballistic
to diffusive regimes is much lowered due to critical fluctuations,
compared with the disordered Fermi liquid
\cite{Paul_Disorder_FMQCP}. Generally speaking, the stability of
the quantum critical point should be addressed, given by the
Harris criterion \cite{Harris}. When the Harris criterion is not
satisfied, three possibilities are expected to arise
\cite{Disorder_QCP_Review2}. The first two possibilities are
emergence of new fixed points, associated with either a
finite-randomness fixed point satisfying the Harris criterion at
this new fixed point or an infinite randomness fixed point
exhibiting activated scaling behaviors. The last possibility is
that quantum criticality can be destroyed, replaced with a smooth
crossover. In addition, even away from the quantum critical point
the disordered system may show non-universal power law physics,
called the Griffiths phase \cite{Griffiths}. Effects of rare
regions are expected to be strong near the infinite randomness
fixed point and the disorder-driven crossover region
\cite{Disorder_QCP_Review2}.

This study focuses on the role of strong randomness in the heavy
fermion quantum transition. Heavy fermion quantum criticality is
believed to result from competition between Kondo and RKKY
(Ruderman-Kittel-Kasuya-Yosida) interactions, where larger Kondo
couplings give rise to a heavy fermion Fermi liquid while larger
RKKY interactions cause an antiferromagnetic metal
\cite{Disorder_QCP_Review2,QCP_Review,HF_Review}. Generally
speaking, there are two competing view points for this problem.
The first direction is to regard the heavy fermion transition as
an antiferromagnetic transition, where critical spin fluctuations
appear from heavy fermions. The second view point is that the
transition is identified with breakdown of the Kondo effect, where
Fermi surface fluctuations are critical excitations. The first
scenario is described by the Hertz-Moriya-Millis (HMM) theory in
terms of heavy electrons coupled with antiferromagnetic spin
fluctuations, the standard model for quantum criticality
\cite{HMM}. There are two ways to realize the second scenario
depending on how to describe Fermi surface fluctuations. The first
way is to express Fermi surface fluctuations in terms of a
hybridization order parameter called holon in the slave-boson
context \cite{KB_z2,KB_z3}. This is usually referred as the Kondo
breakdown scenario. The second one is to map the lattice problem
into the single site one resorting to the dynamical mean-field
theory (DMFT) approximation \cite{DMFT_Review}, where order
parameter fluctuations are critical only in the time direction.
This description is called the locally critical scenario
\cite{EDMFT}.

Each scenario predicts its own critical physics. Both the HMM
theory and the Kondo breakdown model are based on the standard
picture that quantum criticality arises from long-wave-length
critical fluctuations while the locally quantum critical scenario
has its special structure, that is, locally (space) critical
(time). Critical fluctuations are described by $z = 2$ in the HMM
theory due to finite-wave vector ordering \cite{HMM} while by $z =
3$ in the Kondo breakdown scenario associated with uniform
"ordering" \cite{KB_z3}, where $z$ is the dynamical exponent
expressing the dispersion relation for critical excitations. Thus,
quantum critical physics characterized by scaling exponents is
completely different between these two models. In addition to
qualitative agreements with experiments depending on compounds
\cite{Disorder_QCP_Review2}, these two theories do not allow the
$\omega/T$ scaling in the dynamic susceptibility of their critical
modes because both theories live above their upper critical
dimensions. On the other hand, the locally critical scenario gives
rise to the $\omega/T$ scaling behavior for the dynamic spin
susceptibility \cite{EDMFT} while it seems to have some
difficulties associated with some predictions for transport
coefficients.

We start to discuss an Ising model with Gaussian randomness for
its exchange coupling, called the Edwards-Anderson model
\cite{SG_Review}. Using the replica trick and performing the
saddle-point analysis, one can find a spin glass phase when the
average value of the exchange interaction vanishes, characterized
by the Edwards-Anderson order parameter without magnetization.
Applying this concept to the Heisenberg model with Gaussian
randomness, quantum fluctuations should be incorporated to take
into account the Berry phase contribution carefully. It was
demonstrated that quantum corrections in the DMFT approximation
lead the spin glass phase unstable at finite temperatures,
resulting in a spin liquid state when the average value of the
exchange coupling vanishes \cite{Sachdev_SG}. It should be noted
that this spin liquid state differs from the spin liquid phase in
frustrated spin systems in the respect that the former state
originates from critical single-impurity dynamics while the latter
phase results from non-trivial spatial spin correlations described
by gauge fluctuations \cite{Spin_Liquid_Review}. The spin liquid
phase driven by strong randomness is characterized by its critical
spin spectrum, given by the $\omega/T$ scaling local spin
susceptibility \cite{Sachdev_SG}.

Introducing hole doping into the spin liquid state, Parcollet and
Georges examined the disordered t-J model within the DMFT
approximation \cite{Olivier}. Using the U(1) slave-boson
representation, they found marginal Fermi-liquid phenomenology,
where the electrical transport is described by the $T$-linear
resistivity, resulting from the marginal Fermi-liquid spectrum for
collective modes, here the $\omega/T$ scaling in the local spin
susceptibility. They tried to connect this result with physics of
high T$_{c}$ cuprates.

In this study we introduce random hybridization with conduction
electrons into the spin liquid state. Our original motivation was
to explain both the $\omega/T$ scaling in the spin spectrum
\cite{INS_Local_AF} and the typical $T$-linear resistivity
\cite{LGW_F_QPT_Nature} near the heavy fermion quantum critical
point. In particular, the presence of disorder leads us to the
DMFT approximation naturally \cite{Moore_Dis_DMFT}, expected to
result in the $\omega/T$ scaling for the spin spectrum
\cite{Sachdev_SG}.

Starting from an Anderson lattice model with disorder, we derive
an effective local field theory in the DMFT approximation, where
randomness is introduced into both hybridization and RKKY
interactions. Performing the saddle-point analysis in the U(1)
slave-boson representation, we reveal its phase diagram which
shows a quantum phase transition from a spin liquid state to a
local Fermi liquid phase. In contrast with the clean limit of the
Anderson lattice model \cite{KB_z2,KB_z3}, the effective
hybridization given by holon condensation turns out to vanish,
resulting from the zero mean value of the hybridization coupling
constant. However, we show that the holon density becomes finite
when variance of hybridization is sufficiently larger than that of
the RKKY coupling, giving rise to the Kondo effect. On the other
hand, when the variance of hybridization becomes smaller than that
of the RKKY coupling, the Kondo effect disappears, resulting in a
fully symmetric paramagnetic state, adiabatically connected with
the spin liquid state of the disordered Heisenberg model
\cite{Sachdev_SG}.

Our contribution compared with the previous works
\cite{Kondo_Disorder} is to introduce RKKY interactions between
localized spins and to observe the quantum phase transition in the
heavy fermion system with strong randomness. The previous works
focused on how the non-Fermi liquid physics can appear in the
Kondo singlet phase away from quantum criticality
\cite{Kondo_Disorder}. A huge distribution of the Kondo
temperature $T_{K}$ turns out to cause such non-Fermi liquid
physics, originating from the finite density of unscreened local
moments with almost vanishing $T_K$, where the $T_{K}$
distribution may result from either the Kondo disorder for
localized electrons or the proximity of the Anderson localization
for conduction electrons. Because RKKY interactions are not
introduced in these studies, there always exist finite $T_{K}$
contributions. On the other hand, the presence of RKKY
interactions gives rise to breakdown of the Kondo effect, making
$T_{K} = 0$ identically in the strong RKKY coupling phase.

In Ref. [\onlinecite{Kondo_RKKY_Disorder}] the role of random RKKY
interactions was examined, where the Kondo coupling is fixed while
the chemical potential for conduction electrons is introduced as a
random variable with its variance $W$.
%
%
Increasing the randomness of the electron chemical potential, the
Fermi liquid state in $W < W_{c}$ turns into the spin liquid phase
in $W > W_{c}$, which displays the marginal Fermi-liquid
phenomenology due to random RKKY interactions
\cite{Kondo_RKKY_Disorder}, where the Kondo effect is suppressed
due to the proximity of the Anderson localization for conduction
electrons \cite{Kondo_Disorder}. However, the presence of finite
Kondo couplings still gives rise to Kondo screening although the
$T_{K}$ distribution differs from that in the Fermi liquid state,
associated with the presence of random RKKY interactions. In
addition, the spin liquid state was argued to be unstable against
the spin glass phase at low temperatures, maybe resulting from the
fixed Kondo interaction. On the other hand, we do not take into
account the Anderson localization for conduction electrons, and
introduce random hybridization couplings. As a result, the Kondo
effect is completely destroyed in the spin liquid phase, thus
quantum critical physics differs from the previous study of Ref.
[\onlinecite{Kondo_RKKY_Disorder}]. In addition, the spin liquid
phase is stable at finite temperatures in the present study
\cite{Sachdev_SG}.

We investigate the quantum critical point beyond the mean-field
approximation. Introducing quantum corrections fully
self-consistently in the non-crossing approximation
\cite{Hewson_Book}, we prove that the local charge susceptibility
has exactly the same critical exponent as the local spin
susceptibility. This is quite unusual because these correlation
functions are symmetry-unrelated in the lattice scale. This
reminds us of deconfined quantum criticality \cite{Senthil_DQCP},
where the Landau-Ginzburg-Wilson forbidden continuous transition
may appear with an enhanced emergent symmetry. Actually, the
continuous quantum transition was proposed between the
antiferromagnetic phase and the valence bond solid state
\cite{Senthil_DQCP}. In the vicinity of the quantum critical point
the spin-spin correlation function of the antiferromagnetic
channel has the same scaling exponent as the valence-bond
correlation function, suggesting an emergent O(5) symmetry beyond
the symmetry O(3)$\times$Z$_{4}$ of the lattice model
\cite{Tanaka_SO5} and confirmed by the Monte Carlo simulation of
the extended Heisenberg model \cite{Sandvik}. Tanaka and Hu
proposed an effective O(5) nonlinear $\sigma$ model with the
Wess-Zumino-Witten term as an effective field theory for the
Landau-Ginzburg-Wilson forbidden quantum critical point
\cite{Tanaka_SO5}, expected to allow fractionalized spin
excitations due to the topological term. This proposal can be
considered as generalization of an antiferromagnetic spin chain,
where an effective field theory is given by an O(4) nonlinear
$\sigma$ model with the Wess-Zumino-Witten term, which gives rise
to fractionalized spin excitations called spinons, identified with
topological solitons \cite{Tsvelik_Book}. Applying this concept to
the present quantum critical point, the enhanced emergent symmetry
between charge (holon) and spin (spinons) local modes leads us to
propose novel duality between the Kondo singlet phase and the
critical local moment state beyond the Landau-Ginzburg-Wilson
paradigm. We suggest an O(4) nonlinear $\sigma$ model in a
nontrivial manifold as an effective field theory for this local
quantum critical point, where the local spin and charge densities
form an O(4) vector with a constraint. The symmetry enhancement
serves the mechanism of electron fractionalization in critical
impurity dynamics, where such fractionalized excitations are
identified with topological excitations.

This paper is organized as follows. In section II we introduce an
effective disordered Anderson lattice model and perform the DMFT
approximation with the replica trick. Equation (\ref{DMFT_Action})
is the main result in this section. In section III we perform the
saddle-point analysis based on the slave-boson representation and
obtain the phase diagram showing breakdown of the Kondo effect
driven by the RKKY interaction. We show spectral functions,
self-energies, and local spin susceptibility in the Kondo phase.
Figures (1)-(3) with Eqs. (\ref{Sigma_C_MFT})-(\ref{Sigma_FC_MFT})
and (\ref{Lambda_MFT})-(\ref{Constraint_MFT}) are main results in
this section. In section IV we investigate the nature of the
impurity quantum critical point based on the non-crossing
approximation beyond the previous mean-field analysis. We solve
self-consistent equations analytically and find power-law scaling
solutions. As a result, we uncover the marginal Fermi-liquid
spectrum for the local spin susceptibility. We propose an
effective field theory for the quantum critical point and discuss
the possible relationship with the deconfined quantum critical
point. In section V we summarize our results.

The present study extends our recent publication
\cite{Tien_Kim_PRL}, showing both physical and mathematical
details.

\section{An effective DMFT action from an Anderson lattice model with strong randomness}

We start from an effective Anderson lattice model \bqa H &=& -
\sum_{ij,\sigma} t_{ij} c^{\dagger}_{i\sigma} c_{j\sigma} + E_{d}
\sum_{i\sigma} d^{\dagger}_{i\sigma} d_{i\sigma} \nn &+& \sum_{ij}
J_{ij} \mathbf{S}_{i} \cdot \mathbf{S}_{j} + \sum_{i\sigma} (V_{i}
c^{\dagger}_{i\sigma} d_{i\sigma} + {\rm H.c.}) , \label{ALM} \eqa
where $t_{ij} = \frac{t}{M \sqrt{z}}$ is a hopping integral for
conduction electrons and \bqa && J_{ij} = \frac{J}{\sqrt{z M}}
\varepsilon_{i}\varepsilon_{j} , ~~~~~ V_{i} = \frac{V}{\sqrt{M}}
\varepsilon_{i} \nonumber \eqa are random RKKY and hybridization
coupling constants, respectively. Here, $M$ is the spin degeneracy
and $z$ is the coordination number. Randomness is given by the
Gaussian distribution \bqa \overline{\varepsilon_{i}} = 0 , ~~~~~
\overline{\varepsilon_{i}\varepsilon_{j}} = \delta_{ij} . \eqa

The disorder average can be performed in the replica trick
\cite{SG_Review}. Performing the disorder average in the Gaussian
distribution function, we reach the following expression for the
replicated effective action
\begin{eqnarray}
&& \overline{Z^n} = \int \mathcal{D}c_{i\sigma}^{a}
\mathcal{D}d_{i\sigma}^{a} e^{-\bar{S}_n } , \nn &&
\overline{S}_{n} = \int\limits_{0}^{\beta} d\tau \sum_{ij\sigma a}
c^{\dagger a}_{i\sigma}(\tau) ((\partial_{\tau} - \mu)\delta_{ij}
+ t_{ij}) c^{a}_{j\sigma}(\tau) \nn && + \int\limits_{0}^{\beta}
d\tau \sum_{i\sigma a}d^{\dagger a}_{i\sigma}(\tau)
(\partial_{\tau} + E_d) d^{a}_{i\sigma}(\tau) \nn && -
\frac{J^2}{2 z M} \int\limits_{0}^{\beta} d\tau
\int\limits_{0}^{\beta} d\tau' \sum_{ijab}
\mathbf{S}^{a}_{i}(\tau) \cdot \mathbf{S}^{a}_{j}(\tau) \;\;
\mathbf{S}^{b}_{i}(\tau') \cdot \mathbf{S}^{b}_{j}(\tau') \nn && -
\frac{V^{2}}{2 M} \int\limits_{0}^{\beta} d\tau
\int\limits_{0}^{\beta} d\tau' \sum_{i \sigma \sigma' ab} \big(
c^{\dagger a}_{i\sigma}(\tau) d^{a}_{i\sigma}(\tau) + d^{\dagger
a}_{i\sigma}(\tau) c^{a}_{i\sigma}(\tau)\big) \nn &&
~~~~~~~~~~~~~~~ \times \big( c^{\dagger b}_{i\sigma'}(\tau')
d^{b}_{i\sigma'}(\tau') + d^{\dagger b}_{i\sigma'}(\tau')
c^{b}_{i\sigma'}(\tau')\big) , \label{DALM}
\end{eqnarray}
where $\sigma, \sigma' = 1, ..., M$ is the spin index and $a, b =
1, ..., n$ is the replica index. In appendix A we derive this
replicated action from Eq. (\ref{ALM}).

One may ask the role of randomness of $E_{d}$, generating \bqa &&
- \int_{0}^{\beta} d\tau \int_{0}^{\beta} d\tau'
\sum_{i\sigma\sigma' ab} d^{\dagger a}_{i\sigma}(\tau)
d^{a}_{i\sigma}(\tau) d^{\dagger b}_{i\sigma'}(\tau')
d^{b}_{i\sigma'}(\tau') , \nonumber \eqa where density
fluctuations are involved. This contribution is expected to
support the Kondo effect because such local density fluctuations
help hybridization with conduction electrons. In this paper we fix
$E_{d}$ as a constant value in the Kondo limit, allowed as long as
its variance is not too large to overcome the Kondo limit.

One can introduce randomness in the hopping integral of conduction
electrons. But, this contribution gives rise to the same effect as
the DMFT approximation in the $z\rightarrow \infty$ Bethe lattice
\cite{Olivier}. In this respect randomness in the hopping integral
is naturally introduced into the present DMFT study.

The last disorder contribution can arise from randomness in the
electron chemical potential, expected to cause the Anderson
localization for conduction electrons. Actually, this results in
the metal-insulator transition at the critical disorder strength,
suppressing the Kondo effect in the insulating phase. Previously,
the Griffiths phase for non-Fermi liquid physics has been
attributed to the proximity effect of the Anderson localization
\cite{Kondo_Disorder}. In this work we do not consider the
Anderson localization for conduction electrons.

%
%

We observe that the disorder average neutralizes spatial
correlations except for the hopping term of conduction electrons.
This leads us to the DMFT formulation, resulting in an effective
local action for the strong random Anderson lattice model
\begin{eqnarray}
&& \bar{S}_{n}^{\rm eff} = \int_{0}^{\beta} d\tau \Bigl\{
\sum_{\sigma a} c^{\dagger a}_{\sigma}(\tau) (\partial_{\tau} -
\mu) c^{a}_{\sigma}(\tau) \nn && + \sum_{\sigma a}d^{\dagger
a}_{\sigma}(\tau) (\partial_{\tau} + E_d) d^{a}_{\sigma}(\tau)
\Bigr\} \nn && -\frac{V^2}{2 M} \int_{0}^{\beta} d\tau
\int_{0}^{\beta} d\tau' \sum_{\sigma \sigma' a b} \big[ c^{\dagger
a}_{\sigma}(\tau) d^{a}_{\sigma}(\tau) + d^{\dagger
a}_{\sigma}(\tau) c^{a}_{\sigma}(\tau)\big] \nn &&
~~~~~~~~~~~~~~~~~~~~~~~~~ \times \big[ c^{\dagger
b}_{\sigma'}(\tau') d^{b}_{\sigma'}(\tau') + d^{\dagger
b}_{\sigma'}(\tau') c^{b}_{\sigma'}(\tau')\big] \nn && -
\frac{J^2}{2 M} \int_{0}^{\beta} d\tau \int_{0}^{\beta} d\tau'
\sum_{ab} \sum_{\alpha\beta\gamma\delta} S^{a}_{\alpha\beta}(\tau)
R^{ab}_{\beta\alpha\gamma\delta}(\tau-\tau')
S^{b}_{\delta\gamma}(\tau') \nn && + \frac{t^2}{M^2}
\int_{0}^{\beta} d\tau \int_{0}^{\beta} d\tau' \sum_{ab\sigma}
c^{\dagger a}_{\sigma}(\tau) G^{ab}_{c \;
\sigma\sigma}(\tau-\tau') c^{b}_{\sigma}(\tau' ) ,
\label{DMFT_Action}
\end{eqnarray}
where $G^{ab}_{c \; ij\sigma\sigma}(\tau-\tau')$ is the local
Green's function for conduction electrons and $R^{ab}_{\beta
\alpha \gamma \delta}(\tau-\tau')$ is the local spin
susceptibility for localized spins, given by \bqa G^{ab}_{c \;
ij\sigma\sigma}(\tau-\tau') &=& - \langle T_{\tau} [
c^{a}_{i\sigma}(\tau) c^{\dagger b}_{j\sigma}(\tau') ] \rangle ,
\nn R^{ab}_{\beta \alpha \gamma \delta}(\tau-\tau') &=& \langle
T_{\tau} [S^{a}_{\beta\alpha}(\tau) S^{b}_{\gamma\delta}(\tau')]
\rangle , \label{Local_Green_Functions} \eqa respectively. Eq.
(\ref{DMFT_Action}) with Eq. (\ref{Local_Green_Functions}) serves
a completely self-consistent framework for this problem.
Derivation of Eq. (\ref{DMFT_Action}) from Eq. (\ref{DALM}) is
shown in appendix B.

This effective model has two well known limits, corresponding to
the disordered Heisenberg model \cite{Sachdev_SG} and the
disordered Anderson lattice model without RKKY interactions
\cite{Kondo_Disorder}, respectively. In the former case a spin
liquid state emerges due to strong quantum fluctuations while a
local Fermi liquid phase appears at low temperatures in the latter
case as long as the $T_{K}$ distribution is not so broadened
enough. In this respect it is natural to consider a quantum phase
transition driven by the ratio between variances for the RKKY and
hybridization couplings.

\section{Phase diagram}

\subsection{Slave boson representation and mean field approximation}

We solve the effective DMFT action based on the U(1) slave boson
representation
\begin{eqnarray}
d^{a}_{\sigma} &=& \hat{b}^{\dagger a} f^{a}_{\sigma} , \label{SB_Electron} \\
S_{\sigma\sigma'}^{a} &=& f^{a\dagger}_{\sigma} f_{\sigma'}^{a} -
q_{0}^{a} \delta_{\sigma \sigma'} \label{SB_Spin}
\end{eqnarray}
with the single occupancy constraint $|b^{a}|^2 + \sum_{\sigma}
f^{a}_{\sigma}(\tau) f^{a}_{\sigma}(\tau) = 1$, where $q_{0}^{a} =
\sum_{\sigma}f^{a\dagger}_{\sigma} f_{\sigma}^{a}/M $.

In the mean field approximation we replace the holon operator
$\hat{b}^{a}$ with its expectation value $\langle \hat{b}^{a}
\rangle \equiv b^{a}$. Then, the effective action Eq.
(\ref{DMFT_Action}) becomes
\begin{widetext}
\begin{eqnarray}
&& \bar{S}_{n}^{\rm eff} = \int_{0}^{\beta} d\tau \Bigl\{
\sum_{\sigma a} c^{\dagger a}_{\sigma}(\tau) (\partial_{\tau} -
\mu) c^{a}_{\sigma}(\tau) + \sum_{\sigma a} f^{\dagger
a}_{\sigma}(\tau) (\partial_{\tau} + E_d) f^{a}_{\sigma}(\tau) +
\sum_{a} \lambda^{a} (|b^{a}|^2 + \sum_{\sigma}
f^{a}_{\sigma}(\tau) f^{a}_{\sigma}(\tau)- 1) \Bigr\} \nonumber \\
&& -\frac{V^2}{2 M} \int_{0}^{\beta} d\tau \int_{0}^{\beta} d\tau'
\sum_{\sigma \sigma' a b} \big[ c^{\dagger a}_{\sigma}(\tau)
f^{a}_{\sigma}(\tau) (b^{a})^{*} + b^{a} f^{\dagger
a}_{\sigma}(\tau) c^{a}_{\sigma}(\tau)\big] \big[ c^{\dagger
b}_{\sigma'}(\tau') f^{b}_{\sigma'}(\tau') (b^{b})^{*} + b^{b}
f^{\dagger b}_{\sigma'}(\tau') c^{b}_{\sigma'}(\tau')\big]
\nonumber \\ &&-\frac{J^2}{2 M} \int_{0}^{\beta} d\tau
\int_{0}^{\beta} d\tau' \sum_{ab} \sum_{\alpha\beta\gamma\delta}
\big[f^{\dagger a}_{\alpha}(\tau) f^{a}_{\beta}(\tau) -
q_{\alpha}^{a} \delta_{\alpha\beta} \big]
R^{ab}_{\beta\alpha\gamma\delta}(\tau-\tau') \big[f^{\dagger
b}_{\delta}(\tau') f^{b}_{\gamma}(\tau') - q_{\gamma}^{b}
\delta_{\gamma\delta} \big] \nonumber \\ && + \frac{t^2}{M^2}
\int_{0}^{\beta} d\tau \int_{0}^{\beta} d\tau' \sum_{ab\sigma}
c^{\dagger a}_{\sigma}(\tau) G^{ab}_{\sigma}(\tau-\tau')
c^{b}_{\sigma}(\tau' ) , \label{SB_MFT}
\end{eqnarray}
\end{widetext}
where $\lambda^{a}$ is a lagrange multiplier field to impose the
constraint and $q_{\alpha}^{a} =\langle f^{\dagger a}_{\alpha}
f^{a}_{\alpha} \rangle$.

Taking the $M\rightarrow \infty$ limit, we obtain self-consistent
equations for self-energy corrections,
%
%
\begin{eqnarray}
\Sigma_{c \;\sigma\sigma'}^{\;ab}(\tau) &=& \frac{V^2}{M} G_{f \;
\sigma\sigma'}^{\; a b}(\tau) (b^{a})^{*} b^b + \frac{t^2}{M^2}
\delta_{\sigma\sigma'} G_{c \; \sigma}^{\; a b}(\tau) ,
\\ \Sigma_{f \;\sigma\sigma'}^{\;ab}(\tau) &=& \frac{V^2}{M} G_{c
\; \sigma\sigma'}^{\; a b}(\tau) (b^{b})^{*} b^a \nn &+&
\frac{J^2}{2 M} \sum_{s s'} G_{f \; s s'}^{\; a b}(\tau) [
R^{ab}_{s\sigma \sigma' s'}(\tau) + R^{ba}_{\sigma' s' s
\sigma}(-\tau) ] , \nn \\ \Sigma_{cf \; \sigma\sigma'}^{\;\;
ab}(\tau) &=& - \delta_{ab} \delta_{\sigma\sigma'}\delta(\tau)
\frac{V^2}{M} \sum_{s c} [\langle  f^{\dagger c}_{s} c^{c}_{s}
\rangle  b^c + {\rm c.c.} ] (b^{a})^{*} \nn &+& \frac{V^2}{M}
G_{fc \; \sigma\sigma'}^{\;\; ab}(\tau) (b^a b^b)^{*} , \\
\Sigma_{fc \; \sigma\sigma'}^{\;\; ab}(\tau) &=& - \delta_{ab}
\delta_{\sigma\sigma'}\delta(\tau) \frac{V^2}{M} \sum_{s c}
[\langle  f^{\dagger c}_{s} c^{c}_{s} \rangle  b^c + {\rm c.c.} ]
b^{a} \nn &+& \frac{V^2}{M} G_{cf \; \sigma\sigma'}^{\;\;
ab}(\tau) b^a b^b ,
\end{eqnarray} respectively, where local Green's functions are given by
\begin{eqnarray}
G_{c \; \sigma\sigma'}^{\; ab}(\tau) &=& - \langle T_c
c^{a}_{\sigma}(\tau) c^{\dagger b}_{\sigma'} (0) \rangle ,
\\
G_{f \; \sigma\sigma'}^{\; ab}(\tau) &=& - \langle T_c
f^{a}_{\sigma}(\tau) f^{\dagger b}_{\sigma'} (0) \rangle ,
\\
G_{cf \; \sigma\sigma'}^{\; ab}(\tau) &=& - \langle T_c
c^{a}_{\sigma}(\tau) f^{\dagger b}_{\sigma'} (0) \rangle ,
\\
G_{fc \; \sigma\sigma'}^{\; ab}(\tau) &=& - \langle T_c
f^{a}_{\sigma}(\tau) c^{\dagger b}_{\sigma'} (0) \rangle .
\end{eqnarray}

In the paramagnetic and symmetric replica phase these Green's
functions are diagonal in the spin and replica indices, i.e.,
$G^{ab}_{x \sigma\sigma'}(\tau)=\delta_{ab}\delta_{\sigma\sigma'}
G_{x}(\tau)$ with $x=c,f,cf,fc$. Then, we obtain the Dyson
equation
\begin{widetext}
\begin{eqnarray}
\left(\begin{array}{cc} G_{c}(i \omega_l) & G_{fc}(i \omega_l) \\
G_{cf}(i \omega_l) & G_{f}(i \omega_l)
\end{array} \right) = \left( \begin{array}{cc}
i\omega_l + \mu - \Sigma_{c}(i \omega_l) & - \Sigma_{cf}(i
\omega_l) \\
- \Sigma_{fc}(i \omega_l) & i\omega_l - E_d -\lambda -
\Sigma_{f}(i \omega_l)
\end{array} \right)^{-1} ,
\end{eqnarray}
\end{widetext}
where $\omega_l=(2 l+1) \pi T$ with $l$ integer. Accordingly, Eqs.
(9)-(12) are simplified as follows
\begin{eqnarray}
\Sigma_{c}(i\omega_l) &=& \frac{V^2}{M} G_{f}(i\omega_l) |b|^2 +
\frac{t^2}{M^2} G_{c}(i\omega_l) , \label{Sigma_C_MFT} \\
\Sigma_{f}(i\omega_l) &=& \frac{V^2}{M} G_{c}(i\omega_l) |b|^2 +
\frac{J^2}{2 M} T \sum_{s} \sum_{\nu_m} G_{f}(i\omega_l-\nu_m) \nn
&\times& [R_{s\sigma\sigma s}(i\nu_m) + R_{\sigma s
s\sigma}(-i\nu_m) ] , \label{Sigma_F_MFT} \\
\Sigma_{cf}(i\omega_l) &=& \frac{V^2}{M} G_{fc}(i\omega_l)
(b^2)^{*} - n \frac{V^2}{M} (b^2)^{*} \sum_s  \langle
f^{\dagger}_{s} c_{s}
+ c^{\dagger}_{s} f_{s} \rangle , \label{Sigma_CF_MFT} \nn \\
\Sigma_{fc}(i\omega_l) &=& \frac{V^2}{M} G_{cf}(i\omega_l) b^2 - n
\frac{V^2}{M} b^2 \sum_s \langle f^{\dagger}_{s} c_{s} +
c^{\dagger}_{s} f_{s} \rangle   \label{Sigma_FC_MFT}
\end{eqnarray} in the frequency space.
Note that $n$ is the replica index and the last terms in
Eqs.~(\ref{Sigma_CF_MFT})-(\ref{Sigma_FC_MFT}) vanish in the limit
of $n \rightarrow 0$. $R_{s\sigma\sigma s}(i\nu_m)$ is the local
spin susceptibility, given by
\begin{eqnarray}
R_{\sigma s s \sigma}(\tau) = - G_{f \sigma}(-\tau) G_{f s}(\tau)
\label{Spin_Corr_MFT}
\end{eqnarray} in the Fourier transformation.

The self-consistent equation for boson condensation is
\begin{eqnarray}
&& b \Big[ \lambda + 2 V^2 T \sum_{\omega_l} G_{c}(i\omega_l)
G_{f}(i\omega_l) \nn && + V^2 T \sum_{\omega_l} \Bigl\{
G_{fc}(i\omega_l) G_{fc}(i\omega_l) + G_{cf}(i\omega_l)
G_{cf}(i\omega_l)\Bigr\} \Big] =0 . \label{Lambda_MFT} \nn
\end{eqnarray}
The constraint equation is given by
\begin{eqnarray}
|b|^2 + \sum_{\sigma} \langle f^{\dagger}_{\sigma} f_{\sigma}
\rangle = 1 . \label{Constraint_MFT}
\end{eqnarray}

The main difference between the clean and disordered cases is that
the off diagonal Green's function $G_{fc}(i\omega_l)$ should
vanish in the presence of randomness in $V$ with its zero mean
value while it is proportional to the condensation $b$ when the
average value of $V$ is finite. In the present situation we find
$b^{a} = \langle f^{a\dagger}_{\sigma} c_{\sigma}^{a} \rangle = 0$
while $(b^{a})^{*}b^{b} = \langle f^{a\dagger}_{\sigma}
c_{\sigma}^{a} c_{\sigma'}^{b\dagger} f_{\sigma'}^{b} \rangle
\equiv |b|^{2} \delta_{ab} \not= 0$. As a result, Eqs.
(\ref{Sigma_CF_MFT}) and (\ref{Sigma_FC_MFT}) are identically
vanishing in both left and right hand sides. This implies that the
Kondo phase is not characterized by the holon condensation but
described by finite density of holons. It is important to notice
that this gauge invariant order parameter does not cause any kinds
of symmetry breaking for the Kondo effect as it should be.
%
%

\subsection{Numerical analysis}

We use an iteration method in order to solve the mean field
equations (\ref{Sigma_C_MFT}), (\ref{Sigma_F_MFT}),
(\ref{Sigma_CF_MFT}), (\ref{Sigma_FC_MFT}), (\ref{Lambda_MFT}),
and (\ref{Constraint_MFT}). For a given $E_d+\lambda$, we use
iterations to find all Green's functions from Eqs.
(\ref{Sigma_C_MFT})-(\ref{Sigma_FC_MFT}) with Eq.
(\ref{Spin_Corr_MFT}) and $b^2$ from Eq.~(\ref{Lambda_MFT}). Then,
we use Eq.~(\ref{Spin_Corr_MFT}) to calculate $\lambda$ and $E_d$.
We adjust the value of $E_d+\lambda$ in order to obtain the
desirable value for $E_d$. Using the obtained $\lambda$ and $b^2$,
we calculate the Green's functions in the real frequency by
iterations. In the real frequency calculation we introduce the
following functions \cite{Saso}
\begin{eqnarray}
\alpha_{\pm}(t)=\int_{-\infty}^{\infty} d\omega e^{-i \omega t}
\rho_{f}(\omega) f(\pm \omega/T),
\end{eqnarray}
where $\rho_{f}(\omega) = -  {\rm Im} G_{f}(\omega+i0^{+})/\pi$ is
the density of states for f-electrons, and $f(x)=1/(\exp(x)+1)$ is
the Fermi-Dirac distribution function. Then, the self-energy
correction from spin correlations is expressed as follows
\begin{eqnarray}
&& \Sigma_{J}(i\omega_l) \equiv \frac{J^2}{2 M} T \sum_{s}
\sum_{\nu_m} G_{f}(i\omega_l-\nu_m) \nn && ~~~~~~~~~~ \times
[R_{s\sigma\sigma s}(i\nu_m) + R_{\sigma s s\sigma}(-i\nu_m) ] \nn
&& = - i J^2 \int_{0}^{\infty} d t e^{i\omega t} \Bigl( [
\alpha_{+}(t)]^2 \alpha_{-}^{*}(t) + [ \alpha_{-}(t)]^2
\alpha_{+}^{*}(t) \Bigr) . \nn
\end{eqnarray} Performing the Fourier transformation, we
calculate $\alpha_{\pm}(t)$ and obtain $\Sigma_{J}(\omega)$.

\begin{figure}[h]
\includegraphics[width=0.48\textwidth]{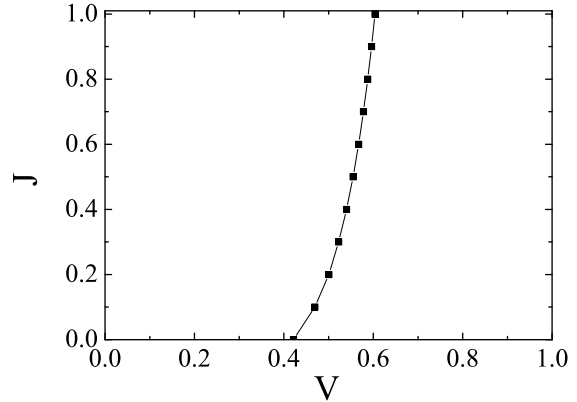}
\caption{The phase diagram of the strongly disordered Anderson
lattice model in the DMFT approximation ($E_d=-1$, $\mu=0$,
$T=0.01$, $t=1$, $M=2$).} \label{fig1}
\end{figure}

\begin{figure}[h]
\includegraphics[width=0.48\textwidth]{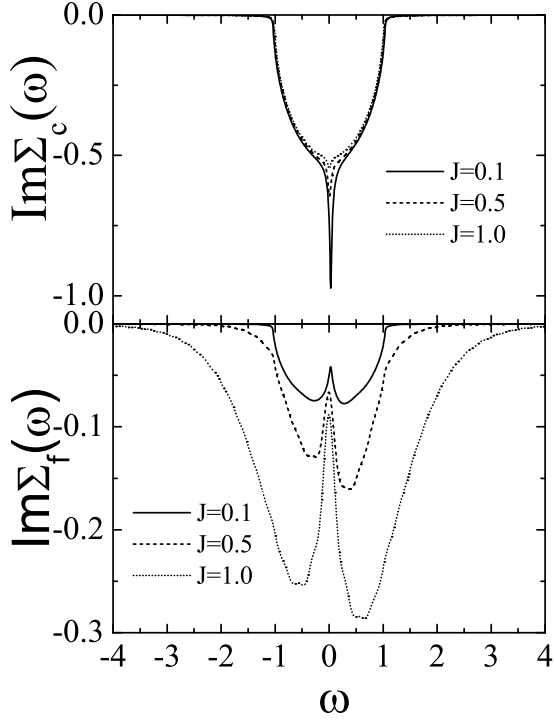}
\caption{The imaginary part of the self-energy of conduction
electrons and that of localized electrons for various values of
$J$ ($V=0.5$, $E_d=-0.7$, $\mu=0$, $T=0.01$, $t=1$, $M$=2).}
\label{fig2}
\end{figure}

\begin{figure}[h]
\includegraphics[width=0.48\textwidth]{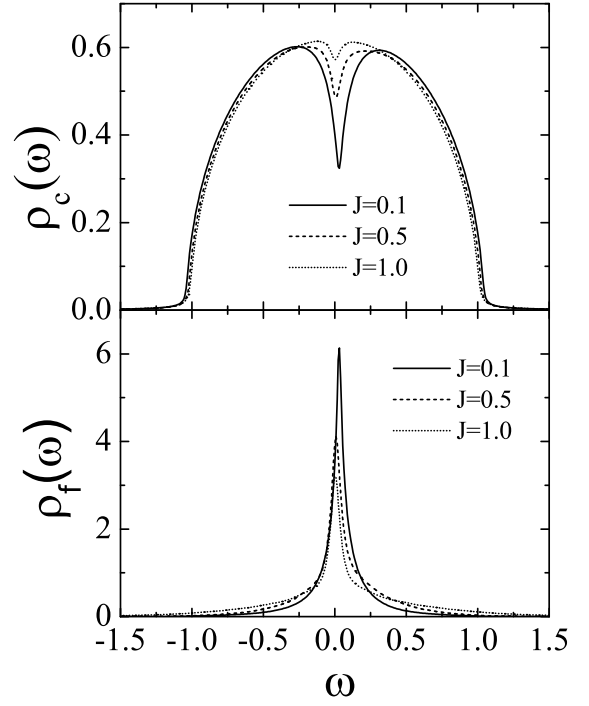}
\caption{Density of states of conduction ($\rho_{c}(\omega)$) and
localized ($\rho_{f}(\omega)$) electrons for various values of $J$
($V=0.5$, $E_d=-0.7$, $\mu=0$, $T=0.01$, $t=1$, $M=2$). }
\label{fig3}
\end{figure}

\begin{figure}[h]
\includegraphics[width=0.48\textwidth]{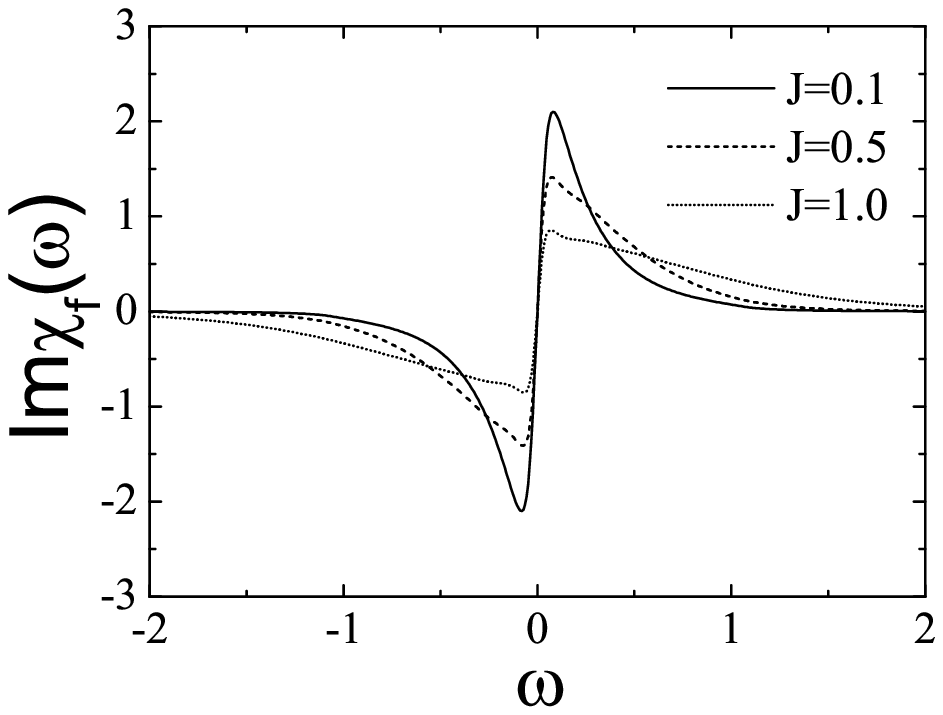}
\caption{Local spin susceptibility for various values of $J$
($V=0.5$, $E_d=-0.7$, $\mu=0$, $T=0.01$, $t=1$, $M=2$).}
\label{fig4}
\end{figure}

Figure \ref{fig1} shows the phase diagram of the strongly
disordered Anderson lattice model in the plane of $(V, J)$, where
$V$ and $J$ are variances for the Kondo and RKKY interactions,
respectively. The phase boundary is characterized by $|b|^{2} =
0$, below which $|b|^{2} \not= 0$ appears to cause effective
hybridization between conduction electrons and localized fermions
although our numerical analysis shows $\langle
f^{\dagger}_{\sigma} c_{\sigma} \rangle =0$, meaning
$\Sigma_{cf(fc)}(i\omega) = 0$ and $G_{cf(fc)}(i\omega) = 0$ in
Eqs. (\ref{Sigma_CF_MFT}) and (\ref{Sigma_FC_MFT}).
%
%

In Fig. \ref{fig2} one finds that the effective hybridization
enhances the scattering rate of conduction electrons dramatically
around the Fermi energy while the scattering rate for localized
electrons becomes reduced at the resonance energy. Enhancement of
the imaginary part of the conduction-electron self-energy results
from the Kondo effect. In the clean situation it is given by the
delta function associated with the Kondo effect
\cite{Hewson_Book}. This self-energy effect reflects the spectral
function, shown in Fig. \ref{fig3}, where the pseudogap feature
arises in conduction electrons while the sharply defined peak
appears in localized electrons, identified with the Kondo
resonance although the description of the Kondo effect differs
from the clean case. Increasing the RKKY coupling, the Kondo
effect is suppressed as expected. In this Kondo phase the local
spin susceptibility is given by Fig. \ref{fig4}, displaying the
typical $\omega$-linear behavior in the low frequency limit,
nothing but the Fermi liquid physics for spin correlations
\cite{Olivier}. Increasing $J$, incoherent spin correlations are
enhanced, consistent with spin liquid physics \cite{Olivier}.

One can check our calculation, considering the $J = 0$ limit to
recover the known result. In this limit we obtain an analytic
expression for $V_c$ at half filling ($\mu=0$)
\begin{eqnarray}
V_c(J=0) &=& \sqrt{\frac{E_d}{2 P_c }}, \\
P_c &=& \int_{-1}^{1} d\omega \rho_{0}(\omega)
\frac{f(\omega/T)-f(0)}{\omega} ,
\end{eqnarray}
where $\rho_{0}(\omega)=\frac{2}{\pi} \sqrt{1-\omega^2}$ is the
bare density of states of conduction electrons. One can check
$V_c(J=0) \rightarrow 0$ in the zero temperature limit because
$P_{c} \rightarrow \infty$.

\section{Nature of quantum criticality}

\subsection{Beyond the saddle-point analysis : Non-crossing approximation}

Resorting to the slave-boson mean-field approximation, we
discussed the phase diagram of the strongly disordered Anderson
lattice model, where a quantum phase transition appears from a
spin liquid state to a dirty "heavy-fermion" Fermi liquid phase,
increasing $V/J$, the ratio of variances of the hybridization and
RKKY interactions. Differentiated from the heavy-fermion quantum
transition in the clean situation, the order parameter turns out
to be the density of holons instead of the holon condensation.
Evaluating self-energies for both conduction electrons and
localized electrons, we could identify the Kondo effect from each
spectral function. In addition, we obtained the local spin
susceptibility consistent with the Fermi liquid physics.

The next task will be on the nature of quantum criticality between
the Kondo and spin liquid phases. This question should be
addressed beyond the saddle-point analysis. Introducing quantum
corrections in the non-crossing approximation, justified in the
$M\rightarrow \infty$ limit, we investigate the quantum critical
point, where density fluctuations of holons are critical.

Releasing the slave-boson mean-field approximation to take into
account holon excitations, we reach the following self-consistent
equations for self-energy corrections,
%
%
\begin{eqnarray}
\Sigma_{c \;\sigma\sigma'}^{\;ab}(\tau) = \frac{V^2}{M} G_{f \;
\sigma\sigma'}^{\; a b}(\tau) G_{b}^{a b}(-\tau) + \frac{t^2}{M^2}
\delta_{\sigma\sigma'} G_{c \; \sigma}^{\; a b}(\tau) ,
\label{Sigma_C_NCA}
\end{eqnarray}
\begin{eqnarray}
\Sigma_{f \;\sigma\sigma'}^{\;ab}(\tau) &=& \frac{V^2}{M} G_{c \;
\sigma\sigma'}^{\; a b}(\tau) G_{b}^{a b}(\tau) \nn &+&
\frac{J^2}{2 M} \sum_{s s'} G_{f \; s s'}^{\; a b}(\tau) [
R^{ab}_{s\sigma \sigma' s'}(\tau) + R^{ba}_{\sigma' s' s
\sigma}(-\tau) ] , \label{Sigma_F_NCA} \nn
\end{eqnarray}
\begin{eqnarray}
\Sigma_{cf \; \sigma\sigma'}^{\;\; ab}(\tau) = - \delta_{ab}
\delta_{\sigma\sigma'}\delta(\tau) \frac{V^2}{M} \sum_{s c} \int
d\tau_1 \langle  f^{\dagger c}_{s} c^{c}_{s} \rangle  G_{b}^{c
a}(\tau_1-\tau') , \label{Sigma_CF_NCA} \nn
\end{eqnarray}
\begin{eqnarray}
\Sigma_{fc \; \sigma\sigma'}^{\;\; ab}(\tau) = - \delta_{ab}
\delta_{\sigma\sigma'}\delta(\tau) \frac{V^2}{M} \sum_{s c}\int
d\tau_1 \langle  c^{\dagger c}_{s} f^{c}_{s} \rangle  G_{b}^{a
c}(\tau-\tau_1) , \label{Sigma_FC_NCA} \nn
\end{eqnarray}
\begin{eqnarray}
\Sigma_{b}^{a b}(\tau) = \frac{V^2}{M} \sum_{\sigma\sigma'} G_{f
\; \sigma\sigma'}^{\; b a}(\tau) G_{c \; \sigma'\sigma}^{\; b
a}(-\tau) . \label{Sigma_B_NCA}
\end{eqnarray}

Since we considered the paramagnetic and replica symmetric phase,
it is natural to assume such symmetries at the quantum critical
point. Note that the off diagonal self-energies,
$\Sigma_{cf}(i\omega_l)$ and $\Sigma_{fc}(i\omega_l)$, are just
constants and proportional to $\langle f^{\dagger}_{\sigma}
c_{\sigma} \rangle$ and $\langle c^{\dagger}_{\sigma} f_{\sigma}
\rangle$, respectively. As a result, $\Sigma_{cf}(i\omega_l) =
\Sigma_{fc}(i\omega_l) = 0$ should be satisfied at the quantum
critical point as the Kondo phase because of $\langle
f^{\dagger}_{\sigma} c_{\sigma} \rangle = \langle
c^{\dagger}_{\sigma} f_{\sigma} \rangle = 0$. Then, we reach the
following self-consistent equations called the non-crossing
approximation
\begin{eqnarray}
\Sigma_{c}(\tau) &=& \frac{V^2}{M} G_{f}(\tau) G_{b}(-\tau) +
\frac{t^2}{M^2} G_{c}(\tau) ,
\label{Sigma_C_NCA_GF} \\
\Sigma_{f}(\tau) &=& \frac{V^2}{M} G_{c}(\tau) G_{b}(\tau) - J^2
[G_{f}(\tau)]^2 G_{f}(-\tau) , \label{Sigma_F_NCA_GF} \\
\Sigma_{b}(\tau) &=& V^2 G_{c}(-\tau) G_{f}(\tau) .
\label{Sigma_B_NCA_GF}
\end{eqnarray}
Local Green's functions are given by
\begin{eqnarray}
G_{c}(i\omega_l) &=& \Big[i\omega_l + \mu - \Sigma_{c}(i\omega_l)
\Big]^{-1} , \label{Dyson_Gc} \\
G_{f}(i\omega_l) &=& \Big[i\omega_l - E_d -\lambda -
\Sigma_{f}(i\omega_l) \Big]^{-1} , \label{Dyson_Gf} \\
G_{b}(i \nu_{l}) &=& \Big[ i\nu_{l} -\lambda -\Sigma_{b}(i\nu_l)
\Big]^{-1} , \label{Dyson_Gb}
\end{eqnarray}
where $\omega_l=(2 l+1) \pi T$ is for fermions and $\nu_{l} = 2 l
\pi T$ is for bosons.

\subsection{Asymptotic behavior at zero temperature }

Calling quantum criticality, power-law scaling solutions are
expected. Actually, if the second term is neglected in Eq.
(\ref{Sigma_F_NCA_GF}), Eqs. (\ref{Sigma_F_NCA_GF}) and
(\ref{Sigma_B_NCA_GF}) are reduced to those of the multi-channel
Kondo effect in the non-crossing approximation \cite{Hewson_Book}.
Power-law solutions are well known in the regime of $1/T_K \ll
\tau \ll \beta=1/T \rightarrow \infty$, where $T_{K} =
D[\Gamma_{c}/\pi D]^{1/M} \exp[\pi E_{d}/M \Gamma_{c}]$ is an
effective Kondo temperature \cite{Tien_Kim} with the conduction
bandwidth $D$ and effective hybridization $\Gamma_{c} = \pi
\rho_{c} \frac{V^{2}}{M}$. In the presence of the RKKY interaction
[the second term in Eq. (\ref{Sigma_F_NCA_GF})], the effective
hybridization will be reduced, where $\Gamma_{c}$ is replaced with
$\Gamma_{c}^{J} \approx \pi \rho_{c} (\frac{V^{2}}{M} - J^{2})$.

Our power-law ansatz is as follows
\begin{eqnarray}
G_{c} &=& \frac{A_c}{\tau^{\Delta_c}} , \\
G_{f} &=& \frac{A_f}{\tau^{\Delta_f}} , \\
G_{b} &=& \frac{A_b}{\tau^{\Delta_b}} ,
\end{eqnarray} where $A_{c}$, $A_{f}$, and $A_{b}$ are positive
numerical constants. In the frequency space these are
\begin{eqnarray}
G_{c}(\omega) &=& A_c C_{\Delta_{c}-1} \omega^{\Delta_c-1}, \label{Dyson_W_Gc} \\
G_{f}(\omega) &=& A_f C_{\Delta_{f}-1} \omega^{\Delta_f-1}, \label{Dyson_W_Gf} \\
G_{b}(\omega) &=& A_b C_{\Delta_{b}-1} \omega^{\Delta_b-1},
\label{Dyson_W_Gb}
\end{eqnarray}
where $C_{\Delta_{c,f,b}} = \int_{-\infty}^{\infty} d x \frac{e^{i
x}}{x^{\Delta_{c,f,b}+1}}.$

Inserting Eqs. (\ref{Dyson_W_Gc})-(\ref{Dyson_W_Gb}) into Eqs.
(\ref{Sigma_C_NCA_GF})-(\ref{Sigma_B_NCA_GF}), we obtain scaling
exponents of $\Delta_{c}$, $\Delta_{f}$, and $\Delta_{b}$. In
appendix C-1 we show how to find such critical exponents in a
detail. Two fixed points are allowed. One coincides with the
multi-channel Kondo effect, given by $\Delta_{c} = 1$, and
$\Delta_{f} = \frac{M}{M+1}$, $\Delta_{b} = \frac{1}{M+1}$ with $M
= 2$, where contributions from spin fluctuations to self-energy
corrections are irrelevant, compared with holon fluctuations. The
other is $\Delta_{c} = 1$ and $\Delta_{f} = \Delta_{b} =
\frac{1}{2}$, where spin correlations are critical as much as
holon fluctuations.

One can understand the critical exponent $\Delta_{f} = 1/2$ as the
proximity of the spin liquid physics \cite{Sachdev_SG}.
Considering the $V \rightarrow 0$ limit, we obtain the scaling
exponents of $\Delta_c = 1$ and $\Delta_f = 1/2$ from the scaling
equations (\ref{92}) and (\ref{93}). Thus, $G_{c}(\omega) \sim
\mbox{sgn}(\omega)$ and $G_{f}(\omega) \sim 1/\sqrt{\omega}$
result for $\omega \rightarrow 0$. In this respect both spin
fluctuations and holon excitations are critical as equal strength
at this quantum critical point.

\subsection{Finite temperature scaling behavior}

We solve Eqs. (\ref{Sigma_C_NCA_GF})-(\ref{Sigma_B_NCA_GF}) in the
regime $\tau, \beta \gg 1/T_K$ with arbitrary $\tau/\beta$, where
the scaling ansatz at zero temperature is generalized as follows
\begin{eqnarray}
G_{c}(\tau) &=& A_{c} \beta^{-\Delta_{c}}
g_{c}\Big(\frac{\tau}{\beta} \Big) , \label{Dyson_T_Gc} \\
G_{f}(\tau) &=& A_{f} \beta^{-\Delta_{f}}
g_{f}\Big(\frac{\tau}{\beta} \Big) , \label{Dyson_T_Gf} \\
G_{b}(\tau) &=& A_{b} \beta^{-\Delta_{b}}
g_{b}\Big(\frac{\tau}{\beta} \Big) . \label{Dyson_T_Gb}
\end{eqnarray}
\begin{eqnarray}
g_{\alpha}(x) = \bigg(\frac{\pi}{\sin(\pi
x)}\bigg)^{\Delta_\alpha} \label{T_Scaling}
\end{eqnarray}
with $\alpha=c,f,b$ is the scaling function at finite
temperatures. In the frequency space we obtain
\begin{eqnarray}
G_{c}(i\omega_l) &=& A_c \beta^{1-\Delta_c}
\Phi_c(i\bar{\omega}_l) , \label{Dyson_TW_Gc} \\
G_{f}(i\omega_l) &=& A_f \beta^{1-\Delta_f}
\Phi_f(i\bar{\omega}_l) , \label{Dyson_TW_Gf} \\
G_{b}(i\nu_l) &=& A_c \beta^{1-\Delta_b} \Phi_b(i\bar{\nu}_l) ,
\label{Dyson_TW_Gb}
\end{eqnarray}
where $\bar{\omega}_l=(2 l+1) \pi$, $\bar{\nu}_l= 2 l \pi$, and
\begin{eqnarray}
\Phi_{\alpha}(i\bar{x}) = \int_{0}^{1} d t e^{i \bar{x} t}
g_{\alpha}(t) . \label{Phi_alpha}
\end{eqnarray}

Inserting Eqs. (\ref{Dyson_TW_Gc})-(\ref{Dyson_TW_Gb}) into Eqs.
(\ref{Sigma_C_NCA_GF})-(\ref{Sigma_B_NCA_GF}), we find two fixed
points, essentially the same as the case of $T = 0$. But, scaling
functions of $\Phi_c(i\bar{\omega}_l)$, $\Phi_f(i\bar{\omega}_l)$,
and $\Phi_b(i\bar{\omega}_l)$ are somewhat complicated. All
scaling functions are derived in appendix C-2.

\subsection{Spin susceptibility}

We evaluate the local spin susceptibility, given by
\begin{eqnarray}
\chi(\tau) &=& G_{f}(\tau) G_{f}(-\tau) , \nonumber \\
&=& A_f^2 \beta^{-2 \Delta_f} \bigg(\frac{\pi}{\sin(\pi
\tau/\beta)} \bigg)^{2\Delta_f} . \label{126}
\end{eqnarray}
The imaginary part of the spin susceptibility
$\chi^{''}(\omega)={\rm Im} \; \chi(\omega+ i0^{+})$ can be found
from
\begin{eqnarray}
\chi(\tau) = \int \frac{d \omega}{\pi} \frac{e^{-\tau
\omega}}{1-e^{-\beta \omega}} \chi^{''}(\omega) . \label{127}
\end{eqnarray}

Inserting the scaling ansatz
\begin{eqnarray}
\chi^{''}(\omega) = A_f^2 \beta^{1-2\Delta_f}
\phi\Big(\frac{\omega}{T}\Big)   \label{128}
\end{eqnarray}
into Eq. (\ref{127}) with Eq. (\ref{126}), we obtain
\begin{eqnarray}
\int \frac{d x}{\pi} \frac{e^{-x \tau/\beta}}{1-e^{-x}} \phi(x) =
\bigg(\frac{\pi}{\sin(\pi \tau/\beta)} \bigg)^{2\Delta_f} .
\end{eqnarray}
Changing the variable $t=i(\tau/\beta -1/2)$, we obtain
\begin{eqnarray}
\int \frac{d x}{\pi} e^{i x t} \frac{\phi(x)}{e^{x}-e^{-x}} =
\bigg(\frac{\pi}{\cosh(\pi t)} \bigg)^{2\Delta_f} .
\end{eqnarray}
As a result, we find the scaling function
\begin{eqnarray}
\phi(x) = 2 (2\pi)^{2 \Delta_f-1} \sinh\Big(\frac{x}{2}\Big)
\frac{\Gamma(\Delta_f+i x/2 \pi)\Gamma(\Delta_f - i
x/2\pi)}{\Gamma(2\Delta_f)} . \nn
\end{eqnarray}
This coincides with the spin spectrum of the spin liquid state
when $V = 0$ \cite{Olivier}.

\subsection{Discussion : Deconfined local quantum criticality}

The local quantum critical point characterized by $\Delta_{c} = 1$
and $\Delta_{f} = \Delta_{b} = 1/2$ is the genuine critical point
in the spin-liquid to local Fermi-liquid transition because such a
fixed point can be connected to the spin liquid state ($\Delta_{c}
= 1$ and $\Delta_{f} = 1/2$) naturally. This fixed point results
from the fact that the spinon self-energy correction from RKKY
spin fluctuations is exactly the same order as that from critical
holon excitations. It is straightforward to see that the critical
exponent of the local spin susceptibility is exactly the same as
that of the local charge susceptibility ($2\Delta_{f} =
2\Delta_{b} = 1$), proportional to $1/\tau$. Since the spinon
spin-density operator differs from the holon charge-density
operator in the respect of symmetry at the lattice scale, the same
critical exponent implies enhancement of the original symmetry at
low energies. The symmetry enhancement sometimes allows a
topological term, which assigns a nontrivial quantum number to a
topological soliton, identified with an excitation of quantum
number fractionalization. This mathematical structure is actually
realized in an antiferromagnetic spin chain \cite{Tsvelik_Book},
generalized into the two dimensional case
\cite{Senthil_DQCP,Tanaka_SO5}.

We propose the following local field theory in terms of physically
observable fields \bqa Z_{eff} &=& \int D
\boldsymbol{\Psi}^{a}(\tau)
\delta\Bigl(|\boldsymbol{\Psi}^{a}(\tau)|^{2} - 1\Bigr) e^{-
\mathcal{S}_{eff}} , \nn \mathcal{S}_{eff} &=& - \frac{g^{2}}{2M}
\int_{0}^{\beta} d \tau \int_{0}^{\beta} d \tau'
\boldsymbol{\Psi}^{a T}(\tau)
\boldsymbol{\Upsilon}^{ab}(\tau-\tau')
\boldsymbol{\Psi}^{b}(\tau') \nn &+& \mathcal{S}_{top} ,
\label{O4_Sigma_Model} \eqa where \bqa &&
\boldsymbol{\Psi}^{a}(\tau) = \left(
\begin{array}{c} \boldsymbol{S}^{a}(\tau) \\ \rho^{a}(\tau)
\end{array} \right) \eqa represents an $O(4)$ vector, satisfying
the constraint of the delta function.
$\boldsymbol{\Upsilon}^{ab}(\tau-\tau')$ determines dynamics of
the $O(4)$ vector, resulting from spin and holon dynamics in
principle. However, it is extremely difficult to derive Eq.
(\ref{O4_Sigma_Model}) from Eq. (\ref{DMFT_Action}) because the
density part for the holon field in Eq. (\ref{O4_Sigma_Model})
cannot result from Eq. (\ref{DMFT_Action}) in a standard way. What
we have shown is that the renormalized dynamics for the O(4)
vector field follows $1/\tau$ asymptotically, where $\tau$ is the
imaginary time. This information should be introduced in
$\boldsymbol{\Upsilon}^{ab}(\tau-\tau')$. $g \propto V/J$ is an
effective coupling constant, and $\mathcal{S}_{top}$ is a possible
topological term.

One can represent the O(4) vector generally as follows
\begin{widetext} \bqa \boldsymbol{\Psi}^{a} : \tau \longrightarrow
\Bigl( \sin \theta^{a}(\tau) \sin \phi^{a}(\tau) \cos
\varphi^{a}(\tau) , \sin \theta^{a}(\tau) \sin \phi^{a}(\tau) \sin
\varphi^{a}(\tau) , \sin \theta^{a}(\tau) \cos \phi^{a}(\tau) ,
\cos \theta^{a}(\tau) \Bigr) , \label{O4_Vector} \eqa
\end{widetext} where $\theta^{a}(\tau), \phi^{a}(\tau),
\varphi^{a}(\tau)$ are three angle coordinates for the O(4)
vector. It is essential to observe that the target manifold for
the O(4) vector is not a simple sphere type, but more complicated
because the last component of the O(4) vector is the charge
density field, where three spin components lie in $- 1 \leq
S^{a}_{x}(\tau), S^{a}_{y}(\tau), S^{a}_{z}(\tau) \leq 1$ while
the charge density should be positive, $0 \leq \rho^{a}(\tau) \leq
1$. This leads us to identify the lower half sphere with the upper
half sphere. Considering that $\sin\theta^{a}(\tau)$ can be folded
on $\pi/2$, we are allowed to construct our target manifold to
have a periodicity, given by
$\boldsymbol{\Psi}^{a}(\theta^{a},\phi^{a},\varphi^{a}) =
\boldsymbol{\Psi}^{a}(\pi - \theta^{a},\phi^{a},\varphi^{a})$.
This folded space allows a nontrivial topological excitation.

Suppose the boundary configuration of
$\boldsymbol{\Psi}^{a}(0,\phi^{a},\varphi^{a}; \tau = 0)$ and
$\boldsymbol{\Psi}^{a}(\pi,\phi^{a},\varphi^{a}; \tau = \beta)$,
connected by $\boldsymbol{\Psi}^{a}(\pi/2,\phi^{a},\varphi^{a}; 0
< \tau < \beta)$. Interestingly, this configuration is {\it
topologically} distinguishable from the configuration of
$\boldsymbol{\Psi}^{a}(0,\phi^{a},\varphi^{a}; \tau = 0)$ and
$\boldsymbol{\Psi}^{a}(0,\phi^{a},\varphi^{a}; \tau = \beta)$ with
$\boldsymbol{\Psi}^{a}(\pi/2,\phi^{a},\varphi^{a}; 0 < \tau <
\beta)$ because of the folded structure. The second configuration
shrinks to a point while the first excitation cannot, identified
with a topologically nontrivial excitation. This topological
excitation carries a spin quantum number $1/2$ in its core, given
by $\boldsymbol{\Psi}^{a}(\pi/2,\phi^{a},\varphi^{a}; 0 < \tau <
\beta) = \Bigl( \sin \phi^{a}(\tau) \cos \varphi^{a}(\tau) , \sin
\phi^{a}(\tau) \sin \varphi^{a}(\tau) , \cos \phi^{a}(\tau) , 0
\Bigr)$. This is the spinon excitation, described by an O(3)
nonlinear $\sigma$ model with the nontrivial spin correlation
function $\boldsymbol{\Upsilon}^{ab}(\tau-\tau')$, where the
topological term is reduced to the single spin Berry phase term in
the instanton core.

In this local impurity picture the local Fermi liquid phase is
described by gapping of instantons while the spin liquid state is
characterized by condensation of instantons. Of course, the low
dimensionality does not allow condensation, resulting in critical
dynamics for spinons. This scenario clarifies the
Landau-Ginzburg-Wilson forbidden duality between the Kondo singlet
and the critical local moment for the impurity state, allowed by
the presence of the topological term.

If the symmetry enhancement does not occur, the effective local
field theory will be given by \bqa Z_{eff} &=& \int
D\boldsymbol{S}^{a}(\tau) D \rho^{a}(\tau) e^{- \mathcal{S}_{eff}}
, \nn \mathcal{S}_{eff} &=& - \int_{0}^{\beta} d \tau
\int_{0}^{\beta} d \tau' \Bigl\{ \frac{V^{2}}{2M} \rho^{a}(\tau)
\chi^{ab}(\tau-\tau') \rho^{b}(\tau') \nn &+& \frac{J^{2}}{2M}
\boldsymbol{S}^{a}(\tau) R^{ab} (\tau-\tau')
\boldsymbol{S}^{b}(\tau') \Bigr\} + \mathcal{S}_{B} \eqa with the
single-spin Berry phase term \bqa \mathcal{S}_{B} = - 2 \pi i S
\int_{0}^{1} d u \int_{0}^{\beta} d \tau \frac{1}{4\pi}
\boldsymbol{S}^{a}(u,\tau)
\partial_{u} \boldsymbol{S}^{a}(u,\tau) \times
\partial_{\tau} \boldsymbol{S}^{a}(u,\tau) , \nonumber \eqa where charge
dynamics $\chi^{ab}(\tau-\tau')$ will be different from spin
dynamics $R^{ab} (\tau-\tau')$. This will not allow the spin
fractionalization for the critical impurity dynamics, where the
instanton construction is not realized due to the absence of the
symmetry enhancement.

\section{Summary}

In this paper we have studied the Anderson lattice model with
strong randomness in both hybridization and RKKY interactions,
where their average values are zero. In the absence of random
hybridization quantum fluctuations in spin dynamics cause the spin
glass phase unstable at finite temperatures, giving rise to the
spin liquid state, characterized by the $\omega/T$ scaling spin
spectrum consistent with the marginal Fermi-liquid phenomenology
\cite{Sachdev_SG}. In the absence of random RKKY interactions the
Kondo effect arises \cite{Kondo_Disorder}, but differentiated from
that in the clean case. The dirty "heavy fermion" phase in the
strongly disordered Kondo coupling is characterized by a finite
density of holons instead of the holon condensation. But,
effective hybridization exists indeed, causing the Kondo resonance
peak in the spectral function. As long as variation of the
effective Kondo temperature is not so large, this disordered Kondo
phase is identified with the local Fermi liquid state because
essential physics results from single impurity dynamics,
differentiated from the clean lattice model.

Taking into account both random hybridization and RKKY
interactions, we find the quantum phase transition from the spin
liquid state to the local Fermi liquid phase at the critical
$(V_{c}, J_{c})$. Each phase turns out to be adiabatically
connected with each limit, i.e., the spin liquid phase when $V =
0$ and the local Fermi liquid phase when $J = 0$, respectively.
Actually, we have checked this physics, considering the local spin
susceptibility and the spectral function for localized electrons.

In order to investigate quantum critical physics, we introduce
quantum corrections from critical holon fluctuations in the
non-crossing approximation beyond the slave-boson mean-field
analysis. We find two kinds of power-law scaling solutions for
self-energy corrections of conduction electrons, spinons, and
holons. The first solution turns out to coincide with that of the
multi-channel Kondo effect, where effects of spin fluctuations are
sub-leading, compared with critical holon fluctuations. In this
respect this quantum critical point is characterized by breakdown
of the Kondo effect while spin fluctuations can be neglected. On
the other hand, the second scaling solution shows that both holon
excitations and spinon fluctuations are critical as the same
strength, reflected in the fact that the density-density
correlation function of holons has the exactly the same critical
exponent as the local spin-spin correlation function of spinons.

We argued that the second quantum critical point implies an
enhanced emergent symmetry from O(3)$\times$O(2)
(spin$\otimes$charge) to O(4) at low energies, forcing us to
construct an O(4) nonlinear $\sigma$ model on the folded target
manifold as an effective field theory for this disorder-driven
local quantum critical point. Our effective local field theory
identifies spinons with instantons, describing the local
Fermi-liquid to spin-liquid transition as the condensation
transition of instantons although dynamics of instantons remains
critical in the spin liquid state instead of condensation due to
low dimensionality. This construction completes novel duality
between the Kondo and critical local moment phases in the strongly
disordered Anderson lattice model.

We explicitly checked that the similar result can be found in the
extended DMFT for the clean Kondo lattice model, where two fixed
point solutions are allowed \cite{EDMFT_Spin,EDMFT_NCA}. One is
the same as the multi-channel Kondo effect and the other is
essentially the same as the second solution in this paper. In this
respect we believe that the present scenario works in the extended
DMFT framework although applicable to only two spatial dimensions
\cite{EDMFT}.

One may suspect the applicability of the DMFT framework for this
disorder problem. However, the hybridization term turns out to be
exactly local in the case of strong randomness while the RKKY term
is safely approximated to be local for the spin liquid state,
expected to be stable against the spin glass phase in the case of
quantum spins. This situation should be distinguished from the
clean case, where the DMFT approximation causes several problems
such as the stability of the spin liquid state \cite{EDMFT_Rosch}
and strong dependence of the dimension of spin dynamics
\cite{EDMFT}.

\section*{Acknowledgement}

This work was supported by the National Research Foundation of
Korea (NRF) grant funded by the Korea government (MEST) (No.
2010-0074542). M.-T. was also supported by the Vietnamese
NAFOSTED.

\appendix

\section{Derivation of Eq. (\ref{DALM}) from Eq. (\ref{ALM}) in the Replica method}

The replica trick \cite{SG_Review} has been utilized for the
disorder average, given by the following identity
\begin{equation}
\overline{\ln Z} = \lim_{n\rightarrow 0}
\frac{\overline{Z^n}-1}{n} ,
\end{equation} where $\overline{\mathcal{O}}$ means the disorder
average of an operator $\mathcal{O}$. $Z^{n}$ is the replicated
partition function
\begin{eqnarray}
Z^{n} = \int \mathcal{D}c_{i\sigma}^{a} \mathcal{D}d_{i\sigma}^{a}
e^{-S_{n}},
\end{eqnarray} where the corresponding replica action is
\begin{eqnarray}
S_{n} && = \int\limits_{0}^{\beta} d\tau \Bigl[ \sum_{i\sigma a}
c^{\dagger a}_{i\sigma}(\tau) (\partial_{\tau} - \mu)
c^{a}_{i\sigma}(\tau) - \sum_{ij\sigma a} t_{ij} c^{\dagger
a}_{i\sigma}(\tau) c^{a}_{j\sigma}(\tau) \nonumber \\ && +
\sum_{i\sigma a} d^{\dagger a}_{i\sigma}(\tau) (\partial_{\tau} +
E_d) d^{a}_{i\sigma}(\tau) + \sum_{ija} J_{ij}
\mathbf{S}^{a}_{i}(\tau) \cdot \mathbf{S}^{a}_{j}(\tau) \nonumber
\\ && + \sum_{i\sigma a} (V_{i} c^{\dagger a}_{i\sigma}(\tau)
d^{a}_{i\sigma}(\tau) + {\rm H.c.}) \Bigr]
\end{eqnarray} with the spin index $\sigma = 1, ..., M$ and the replica index $a = 1, ..., n$.

The disorder average for the replicated partition function is
straightforward, given by
\begin{eqnarray}
\overline{Z^n} = \int d\varepsilon_{i} P[\varepsilon_{i}] \int
\mathcal{D}c_{i\sigma}^{a} \mathcal{D}d_{i\sigma}^{a} e^{-S_{n}} ,
\end{eqnarray} where $P[\varepsilon_{i}]$ is the
Gaussian distribution function with $\int d\varepsilon_{i}
P[\varepsilon_{i}] = 1$. Performing integrals for random
variables, we obtain an effective action Eq. (\ref{DALM}).

\section{Derivation of Eq. (\ref{DMFT_Action}) from Eq. (\ref{DALM}) in the cavity method}

We solve the replicated Anderson lattice model Eq. (\ref{DALM}) in
the DMFT approximation. We apply the cavity method for the DMFT
mapping \cite{DMFT_Review}
\begin{eqnarray}
\overline{S}_{n} =  \bar{S}_{n}^{0} + \Delta \bar{S}_{n} +
\bar{S}_{n}^{(0)} ,
\end{eqnarray}
where $\bar{S}_{n}^{0}$ is the part of the action at a particular
site $0$, $\Delta \bar{S}_{n}$ is the part of the action
connecting the site $0$ with other sites, given by
\begin{widetext}
\begin{eqnarray}
&& \bar{S}_{n}^{0} = \int_{0}^{\beta} d\tau \sum_{\sigma a}
\Bigl\{ c^{\dagger a}_{0\sigma}(\tau) (\partial_{\tau} - \mu)
c^{a}_{0\sigma}(\tau) + d^{\dagger a}_{0\sigma}(\tau)
(\partial_{\tau} + E_d) d^{a}_{0\sigma}(\tau) \Bigr\} \nonumber \\
&& - \frac{V^2}{2 M} \int_{0}^{\beta} d\tau \int_{0}^{\beta}
d\tau' \sum_{\sigma \sigma' ab} \big[ c^{\dagger
a}_{0\sigma}(\tau) d^{a}_{0\sigma}(\tau) + d^{\dagger
a}_{0\sigma}(\tau) c^{a}_{0\sigma}(\tau)\big] \big[ c^{\dagger
b}_{0\sigma'}(\tau') d^{b}_{0\sigma'}(\tau') + d^{\dagger
b}_{0\sigma'}(\tau') c^{b}_{0\sigma'}(\tau')\big] ,
\end{eqnarray}
\end{widetext}
\begin{eqnarray}
&& \Delta \bar{S}_{n} = - \int_{0}^{\beta} d\tau \sum_{i\sigma a}
\Bigl\{ c^{\dagger a}_{0\sigma}(\tau) t_{0i} c^{a}_{i\sigma}(\tau)
+ c^{\dagger a}_{i\sigma}(\tau) t_{i0} c^{a}_{0\sigma}(\tau)
\Bigr\} \nonumber \\ && -\frac{J^2}{2 z M} \int_{0}^{\beta}d\tau
\int_{0}^{\beta} d\tau' \sum_{i a b} \mathbf{S}^{a}_{0}(\tau)
\cdot \mathbf{S}^{a}_{i}(\tau) \;\; \mathbf{S}^{b}_{0}(\tau')
\cdot \mathbf{S}^{b}_{i}(\tau') , \nn
\end{eqnarray} respectively, and $\bar{S}_{n}^{(0)}$
is the rest part of the action.

The partition function can be expanded as follows
\begin{eqnarray}
\overline{Z^{n}} &=& \int \mathcal{D}c_{i\sigma}^{a}
\mathcal{D}d_{i\sigma}^{a} \exp(-\bar{S}_n^{0} - \bar{S}_{n}^{(0)}
- \Delta \bar{S}_{n} ) \nn &=& \int \mathcal{D}c^{a}_{0\sigma}
\mathcal{D}d^{a}_{0\sigma}
\frac{e^{-\bar{S}_{n}^{0}}}{\bar{Z^{n}}^{(0)}} \Big[ 1 -
\int_{0}^{\beta} d\tau \langle \Delta \mathcal{L}(\tau)
\rangle^{(0)} \nn &+& \frac{1}{2!} \int_{0}^{\beta} d\tau
\int_{0}^{\beta} d\tau'
 \langle T_{\tau} \Delta \mathcal{L}(\tau) \Delta \mathcal{L}(\tau')
\rangle^{(0)} + \cdots \Big], \label{Cavity_Expansion}
\end{eqnarray}
where \bqa && \langle \mathcal{O} \rangle^{(0)} \equiv
\frac{1}{\bar{Z^{n}}^{(0)}} \int \mathcal{D}c_{i\not= 0\sigma}^{a}
\mathcal{D}d_{i\not= 0\sigma}^{a} \mathcal{O} \exp( -
\bar{S}_{n}^{(0)} ) \nonumber \eqa with $\bar{Z^{n}}^{(0)} = \int
\mathcal{D}c_{i\not= 0\sigma}^{a} \mathcal{D}d_{i\not=
0\sigma}^{a} \exp( - \bar{S}_{n}^{(0)} )$ and $\Delta \bar{S}_{n}
= \int_{0}^{\beta} d\tau \Delta \mathcal{L}(\tau)$.

The non-trivial term in the first order is given by
\begin{widetext}
\begin{eqnarray} \int_{0}^{\beta} d\tau \langle \Delta \mathcal{L}(\tau)
\rangle^{(0)} &=&  -\frac{J^2}{2 z M} \int_{0}^{\beta} d\tau
\int_{0}^{\beta} d\tau' \sum_{i a b} \langle T_\tau
\mathbf{S}^{a}_{0}(\tau) \cdot \mathbf{S}^{a}_{i}(\tau) \;\;
\mathbf{S}^{b}_{0}(\tau') \cdot \mathbf{S}^{b}_{i}(\tau')
\rangle^{(0)} \nn &=& -\frac{J^2}{2 M} \int_{0}^{\beta} d\tau
\int_{0}^{\beta} d\tau' \sum_{a b} \sum_{\alpha \beta \gamma
\delta} S^{a}_{0 \alpha \beta}(\tau) R^{ab \; (0)}_{\beta \alpha
\gamma \delta}(\tau-\tau') S^{b}_{0 \gamma \delta}(\tau') ,
\end{eqnarray}
\end{widetext}
where $R^{ab \; (0)}_{\beta \alpha \gamma \delta}(\tau-\tau') =
\langle T_{\tau} S^{a}_{\beta\alpha}(\tau)
S^{b}_{\gamma\delta}(\tau')\rangle^{(0)}$. The second order term
is
\begin{widetext}
\begin{eqnarray} \int_{0}^{\beta}d\tau \int_{0}^{\beta} d\tau' \langle T_{\tau}
\Delta \mathcal{L}(\tau) \Delta \mathcal{L}(\tau') \rangle^{(0)}
&=& \int_{0}^{\beta}d\tau \int_{0}^{\beta} d\tau' \sum_{ij\sigma a
b}\langle T_{\tau} c^{\dagger a}_{0\sigma}(\tau) t_{0i}
c^{a}_{i\sigma}(\tau) c^{\dagger b}_{j\sigma}(\tau') t_{j0}
c^{b}_{0\sigma}(\tau' ) \rangle^{(0)} \nn &=& -
\int_{0}^{\beta}d\tau \int_{0}^{\beta} d\tau' \sum_{ij\sigma a b}
c^{\dagger a}_{0\sigma}(\tau) t_{0i} G^{ab \; (0)}_{c \;
ij\sigma\sigma}(\tau-\tau') t_{j0} c^{b}_{0\sigma}(\tau' ) ,
\end{eqnarray}
\end{widetext}
where $G^{ab \; (0)}_{c \; ij\sigma\sigma}(\tau-\tau')= -\langle
T_{\tau} c^{a}_{i\sigma}(\tau) c^{\dagger b}_{j\sigma}(\tau')
\rangle^{(0)}$. One can easily verify that all higher order
expansions in Eq. (\ref{Cavity_Expansion}) vanish in the limit
$z\rightarrow \infty$, which is at the heart of the DMFT
approximation \cite{DMFT_Review}.

In the $z\rightarrow \infty$ Bethe lattice we perform further
simplification\cite{DMFT_Review}
\begin{eqnarray}
&& G^{ab \; (0)}_{c \; ij\sigma\sigma}(\tau) = \delta_{ij}
G^{ab}_{c \; ii\sigma}(\tau) \equiv \delta_{ij} G^{ab}_{c \;
\sigma\sigma}(\tau) , \nn && R^{ab \; (0)}_{\beta \alpha \gamma
\delta}(\tau) = R^{ab}_{\beta \alpha \gamma \delta}(\tau) .
\end{eqnarray}
As a result, we reach an effective single-site action Eq.
(\ref{DMFT_Action}) called the DMFT approximation.

\section{Derivation of critical exponents, $\Delta_{c}$, $\Delta_{f}$, and $\Delta_{b}$}

\subsection{At zero temperature}

Inserting Eqs. (\ref{Dyson_W_Gc})-(\ref{Dyson_W_Gb}) into Eqs.
(\ref{Sigma_C_NCA_GF})-(\ref{Sigma_B_NCA_GF}), we obtain
\begin{eqnarray}
\Sigma_{c}(\omega) &=& \frac{V^2}{M} A_f A_b
C_{\Delta_f+\Delta_b-1} \omega^{\Delta_f+\Delta_b-1} \nn &+&
\frac{t^2}{M^2} A_c C_{\Delta_c-1} \omega^{\Delta_c-1}, \\
\Sigma_{f}(\omega) &=& \frac{V^2}{M} A_c A_b
C_{\Delta_c+\Delta_b-1} \omega^{\Delta_c+\Delta_b-1} \nn &-&
J^2 A_f^3 C_{3\Delta_f-1} \omega^{3\Delta_f-1}, \\
\Sigma_{b}(\omega) &=& V^2 A_c A_f C_{\Delta_c+\Delta_f-1}
\omega^{\Delta_c+\Delta_f-1} .
\end{eqnarray}
It is naturally expected
\begin{eqnarray}
\Sigma_{f}(0^+) &=& - E_d -\lambda , \\
\Sigma_b(0^+) &=& -\lambda
\end{eqnarray} for power-law solutions at zero temperature.

Combined with the Dyson equations
(\ref{Dyson_Gc})-(\ref{Dyson_Gb}), we reach the following
equations
\begin{eqnarray}
\frac{1}{A_c C_{\Delta_c-1}} \omega^{1-\Delta_c} &=& \omega + \mu
- \frac{V^2}{M} A_f A_b C_{\Delta_f+\Delta_b-1}
\omega^{\Delta_f+\Delta_b-1} \nn &-&
\frac{t^2}{M^2} A_c C_{\Delta_c-1} \omega^{\Delta_c-1}, \label{Eq_Gc}\\
\frac{1}{A_f C_{\Delta_f-1}} \omega^{1-\Delta_f} &=&
-\frac{V^2}{M} A_c A_b C_{\Delta_c+\Delta_b-1}
\omega^{\Delta_c+\Delta_b-1} \nn &+&
J^2 A_f^3 C_{3\Delta_f-1} \omega^{3\Delta_f-1}, \label{Eq_Gc}\\
\frac{1}{A_b C_{\Delta_b-1}} \omega^{1-\Delta_b} &=& - V^2 A_c A_f
C_{\Delta_c+\Delta_f-1} \omega^{\Delta_c+\Delta_f-1} .
\label{Eq_Gb}
\end{eqnarray}

The last equation gives
\begin{eqnarray}
\Delta_c+\Delta_f+\Delta_b = 2 .
\end{eqnarray}
Comparing this with the first equation, we get
\begin{eqnarray}
\Delta_c=1 .
\end{eqnarray}

The second equation gives two possible solutions. One is again
$\Delta_c+\Delta_f+\Delta_b = 2$, and the other
$\Delta_f=\Delta_b=1/2$. We can find the first solution by
equating the coefficients in Eqs.~(\ref{Eq_Gc})-(\ref{Eq_Gb}) and
obtain
\begin{eqnarray}
\frac{1}{C_{\Delta_f-1} C_{\Delta_b}} &=& - \frac{V^2}{M} A_c A_f
A_b , \\
\frac{1}{C_{\Delta_b-1} C_{\Delta_f}} &=& - V^2 A_c A_f A_b .
\end{eqnarray}
These two equations result in $$ C_{\Delta_f-1} C_{\Delta_b} = M
C_{\Delta_b-1} C_{\Delta_f}.$$ Using the property of $C_{\Delta-1}
= \Delta C_{\Delta}$, we obtain
$$
\Delta_f = M \Delta_b .
$$
As a result, the first solution is
\begin{eqnarray}
\Delta_f &=& \frac{M}{M+1} , \\
\Delta_b &=& \frac{1}{M+1} ,
\end{eqnarray} exactly the same as those of the multi-channel
Kondo effect.

\subsection{At finite temperatures}

Inserting Eqs. (\ref{Dyson_TW_Gc})-(\ref{Dyson_TW_Gb}) into Eqs.
(\ref{Sigma_C_NCA_GF})-(\ref{Sigma_B_NCA_GF}), we obtain
\begin{eqnarray}
\Sigma_{c}(i\omega_l) &=& \frac{V^2}{M} A_f A_b
\beta^{1-\Delta_f-\Delta_b} \Psi_{fb}(i\bar{\omega}_l) \nn &+&
\frac{t^2}{M^2} \beta^{1-\Delta_c}\Phi_{c}(i\bar{\omega}_l) , \\
\Sigma_{f}(i\omega_l) &=& \frac{V^2}{M} A_c A_b
\beta^{1-\Delta_c-\Delta_b} \Psi_{cb}(i\bar{\omega}_l) \nn &-&
J^2 A_f^3 \beta^{1-3\Delta_f} \Psi_{fff}(i\bar{\omega}_l) , \\
\Sigma_{b}(i\nu_l) &=& V^2 A_c A_f \beta^{1-\Delta_c-\Delta_f}
\Psi_{cf}(i\bar{\nu}_l)  ,
\end{eqnarray}
where
\begin{eqnarray}
\Psi_{fb}(i\bar{\omega}_l) &=& \int_{0}^{1} d t e^{i
\bar{\omega}_l t} g_{f}(t) g_{b}(-t) , \label{Psi_fb} \\
\Psi_{cb}(i\bar{\omega}_l) &=& \int_{0}^{1} d t e^{i
\bar{\omega}_l t} g_{c}(t) g_{b}(t) , \label{Psi_cb} \\
\Psi_{fff}(i\bar{\omega}_l) &=& \int_{0}^{1} d t e^{i
\bar{\omega}_l t} [g_{f}(t)]^2 g_{f}(-t) , \label{Psi_fff} \\
\Psi_{cf}(i\bar{\nu}_l) &=& \int_{0}^{1} d t e^{i \bar{\nu}_l t}
g_{c}(-t) g_{f}(t) . \label{Psi_cf}
\end{eqnarray}

Using the Dyson equations, we obtain the final self-consistency
expressions
\begin{widetext}
\begin{eqnarray}
A_{c}^{-1} \Phi_{c}^{-1}(i\bar{\omega}_l) &=& [i\bar{\omega}_l T +
\mu] \beta^{1-\Delta_c} - \frac{V^2}{M} A_f A_b
\beta^{2-\Delta_c-\Delta_f-\Delta_b} \Psi_{fb}(i\bar{\omega}_l) -
\frac{t^2}{M^2} \beta^{2-2\Delta_c} \Phi_{c}(i\bar{\omega}_l) , \label{92} \\
A_{f}^{-1} \Phi_{f}^{-1}(i\bar{\omega}_l) &=& i\bar{\omega}_l
\beta^{-\Delta_f} -
[E_d+\lambda-\Sigma_{f}(i\omega_{0})]\beta^{1-\Delta_f} -
\frac{V^2}{M} A_c A_b \beta^{2-\Delta_c-\Delta_f-\Delta_b}
[\Psi_{cb}(i\bar{\omega}_l)- \Psi_{cb}(i\bar{\omega}_0)] \nn &+&
J^2 A_f^3 \beta^{2-4\Delta_f}
[\Psi_{fff}(i\bar{\omega}_l)-\Psi_{fff}(i\bar{\omega}_0)] ,
\label{93} \\
A_{b}^{-1} \Phi_{b}^{-1}(i\bar{\nu}_l) &=&i\bar{\nu}_l
\beta^{-\Delta_b} - [\lambda-\Sigma_b(i\bar{\nu}_0)]
\beta^{1-\Delta_b} + V^2 A_c A_f
\beta^{2-\Delta_c-\Delta_f-\Delta_b}
[\Psi_{cf}(i\bar{\nu}_l)-\Psi_{cf}(i\bar{\nu}_0)] . \label{94}
\end{eqnarray}
\end{widetext}

As the zero temperature case, we obtain two power-law solutions,
comparing the powers of $\beta$ terms. One is
\begin{eqnarray}
\Delta_c = 1 , \\
\Delta_f + \Delta_b =1 ,
\end{eqnarray}
with $\Delta_f > 1/2$. The other solution is
\begin{eqnarray}
\Delta_c = 1 , \\
\Delta_f = \Delta_b =1/2 .
\end{eqnarray}
Note that Eqs. (\ref{92}) and (\ref{94}) are same for both
solutions. Only Eq. (\ref{93}) distinguishes these two solutions.

Inserting Eq. (\ref{T_Scaling}) into Eqs. (\ref{Phi_alpha}),
(\ref{Psi_fb})-(\ref{Psi_cb}), we obtain
\begin{widetext}
\begin{eqnarray}
\Phi_{c}(i\bar{\omega}_l) &=& i \pi \; {\rm sgn}(\bar{\omega}_l) , \label{99}\\
\Phi_{f}(i\bar{\omega}_l) &=& (2 \pi)^{\Delta_f} i (-1)^{l}
\frac{\Gamma(1-\Delta_f)}{\Gamma(1-\frac{\Delta_f}{2}+\frac{\bar{\omega}_l}{2\pi})
\Gamma(1-\frac{\Delta_f}{2}-\frac{\bar{\omega}_l}{2\pi})},
\label{100}
\\
\Phi_{b}(i\bar{\nu}_l) &=& (2 \pi)^{\Delta_b} (-1)^{l}
\frac{\Gamma(1-\Delta_b)}{\Gamma(1-\frac{\Delta_b}{2}+\frac{\bar{\nu}_l}{2\pi})
\Gamma(1-\frac{\Delta_b}{2}-\frac{\bar{\nu}_l}{2\pi})} \label{101}
\end{eqnarray} and
\begin{eqnarray}
\Psi_{fb}(i\bar{\omega}_l) &=& i \pi \; {\rm sgn}(\bar{\omega}_l)
, \label{102}\\
\Psi_{cb}(i\bar{\omega}_l) &=& (2 \pi)^{\Delta_b+1} i (-1)^{l}
\frac{\Gamma(-\Delta_b)}{\Gamma(\frac{1}{2}-\frac{\Delta_b}{2}+\frac{\bar{\omega}_l}{2\pi})
\Gamma(\frac{1}{2}-\frac{\Delta_b}{2}-\frac{\bar{\omega}_l}{2\pi})}
, \label{103}\\
\Psi_{fff}(i\bar{\omega}_l) &=& (2 \pi)^{3\Delta_f} i(-1)^{l}
\frac{\Gamma(1-3\Delta_f)}{\Gamma(1-\frac{3\Delta_f}{2}+\frac{\bar{\omega}_l}{2\pi})
\Gamma(1-\frac{3\Delta_f}{2}-\frac{\bar{\omega}_l}{2\pi})}
, \label{104}\\
\Psi_{cf}(i\bar{\nu}_l) &=& (2 \pi)^{\Delta_f+1} (-1)^{l+1}
\frac{\Gamma(-\Delta_f)}{\Gamma(\frac{1}{2}-\frac{\Delta_f}{2}+\frac{\bar{\nu}_l}{2\pi})
\Gamma(\frac{1}{2}-\frac{\Delta_f}{2}-\frac{\bar{\nu}_l}{2\pi})} .
\label{105}
\end{eqnarray}
\end{widetext}

Inserting these expressions into Eq.~(\ref{92}), we obtain the
equation for $A_c$
\begin{eqnarray}
\frac{1}{\pi A_c} + \frac{t^2}{M^2} \pi A_c = i\mu \; {\rm
sgn}(\bar{\omega}_l) + \frac{V^2}{M} \pi  A_f A_b . \label{106}
\end{eqnarray}

From Eq.~(\ref{94}) we obtain the condition
\begin{eqnarray}
\Sigma_b(i\bar{\nu}_{0}) -\lambda &=& T^{1-\Delta_b} \frac{1}{A_b
\Phi_{b}(i\bar{\nu}_0)} , \nonumber \\
&=& \frac{T^{1-\Delta_b}}{A_b (2 \pi)^{\Delta_b}}
\frac{[\Gamma(1-\frac{\Delta_b}{2})]^2}{(-1)^n \Gamma(1-\Delta_b)}
\end{eqnarray}
and the equation
\begin{eqnarray}
A_b^{-1} [\Phi_{b}(i\bar{\nu}_l)-\Phi_{b}(i\bar{\nu}_0)] = V^2 A_c
A_f [\Psi_{cf}(i\bar{\nu}_l)-\Psi_{cf}(i\bar{\nu}_0)]  . \nn
\label{108}
\end{eqnarray}
Inserting Eqs. (\ref{101}) and (\ref{105}) into Eq. (\ref{108}),
we obtain
\begin{widetext}
\begin{eqnarray}
&&
\bigg[\frac{\Gamma(1-\frac{\Delta_b}{2}+\frac{\bar{\nu}_l}{2\pi})
\Gamma(1-\frac{\Delta_b}{2}-\frac{\bar{\nu}_l}{2\pi})}{(-1)^l
\Gamma(1-\Delta_b)} - (l=0) \bigg] = V^2 A_c A_f A_b (2 \pi)^2
\bigg[
\frac{(-1)^{l+1}\Gamma(-\Delta_f)}{\Gamma(\frac{1}{2}-\frac{\Delta_f}{2}+\frac{\bar{\nu}_l}{2\pi})
\Gamma(\frac{1}{2}-\frac{\Delta_f}{2}-\frac{\bar{\nu}_l}{2\pi})}
-(l=0) \bigg] . \label{109}
\end{eqnarray}
One can show that
\begin{eqnarray}
\bigg[\frac{\Gamma(1-\frac{\Delta_b}{2}+\frac{\bar{\nu}_l}{2\pi})
\Gamma(1-\frac{\Delta_b}{2}-\frac{\bar{\nu}_l}{2\pi})}{(-1)^l
\Gamma(1-\Delta_b)} - (l=0) \bigg] =
\frac{[\Gamma(\frac{1}{2}+\frac{\Delta_f}{2})]^2}{\Gamma(1-\Delta_b)}
\Bigg[ \frac{\prod_{k=1}^{l}(k-\frac{1}{2}+\frac{\Delta_f}{2})}
{\prod_{k=1}^{l}(k-\frac{1}{2}-\frac{\Delta_f}{2})} -1\Bigg] , \label{110} \\
\bigg[
\frac{(-1)^{l+1}\Gamma(-\Delta_f)}{\Gamma(\frac{1}{2}-\frac{\Delta_f}{2}+\frac{\bar{\nu}_l}{2\pi})
\Gamma(\frac{1}{2}-\frac{\Delta_f}{2}-\frac{\bar{\nu}_l}{2\pi})}
-(l=0) \bigg] = -
\frac{\Gamma(-\Delta_f)}{[\Gamma(\frac{1}{2}-\frac{\Delta_f}{2})]^2}
\Bigg[ \frac{\prod_{k=1}^{l}(k-\frac{1}{2}+\frac{\Delta_f}{2})}
{\prod_{k=1}^{l}(k-\frac{1}{2}-\frac{\Delta_f}{2})} -1\Bigg] .
\label{111}
\end{eqnarray}
\end{widetext}
Here we have used $\Delta_f + \Delta_b =1$. From
Eqs.~(\ref{109})-(\ref{111}) we obtain
\begin{eqnarray}
1&=& - V^2 A_c A_f A_b (2 \pi)^2 \frac{\Gamma(-\Delta_f)
\Gamma(1-\Delta_b)}{\Big[\Gamma(\frac{1}{2}+\frac{\Delta_f}{2})
\Gamma(\frac{1}{2}-\frac{\Delta_f}{2})\Big]^2} \nonumber \\
&=& - V^2 A_c A_f A_b \Gamma(-\Delta_f) \Gamma(1-\Delta_b) [2
\sin(\pi \Delta_f /2)]^2 . \label{112} \nn
\end{eqnarray}

Equation (\ref{93}) gives the condition
\begin{eqnarray}
 \Sigma_f(i\bar{\omega}_0)- E_d - \lambda &=&  T^{1-\Delta_f}
\frac{1}{A_f \Phi_f(i\bar{\omega}_0)}  \nonumber \\
&=&  \frac{T^{1-\Delta_f}}{A_f (2 \pi)^{\Delta_f}}
\frac{\Gamma(\frac{3}{2}-\frac{\Delta_f}{2})
\Gamma(\frac{1}{2}-\frac{\Delta_f}{2})}{i (-1)^l
\Gamma(1-\Delta_f)}. \nn
\end{eqnarray}
The scaling equation (\ref{93}) distinguishes two solutions, and
we will consider them separately.

\subsubsection{First solution: $\Delta_f+\Delta_b=1$ with $\Delta_f > 1/2$}

The scaling Eq. (\ref{93}) becomes
\begin{eqnarray}
A_f^{-1} [\Phi_{f}(i\bar{\omega}_l)-\Phi_{f}(i\bar{\omega}_0)] =
-\frac{V^2}{M} A_c A_b [\Psi_{cb}(i\bar{\omega}_l)-
\Psi_{cb}(i\bar{\omega}_0)] . \label{114} \nn
\end{eqnarray}
Inserting Eqs. (\ref{100}), (\ref{103}) into Eq. (\ref{114}), we
obtain
\begin{widetext}
\begin{eqnarray}
&&
\bigg[\frac{\Gamma(1-\frac{\Delta_f}{2}+\frac{\bar{\omega}_l}{2\pi})
\Gamma(1-\frac{\Delta_f}{2}-\frac{\bar{\omega}_l}{2\pi})}{(-1)^l
\Gamma(1-\Delta_f)} - (l=0) \bigg] = \frac{V^2}{M} A_c A_f A_b (2
\pi)^2 \bigg[
\frac{(-1)^{l}\Gamma(-\Delta_b)}{\Gamma(\frac{1}{2}-\frac{\Delta_b}{2}+\frac{\bar{\omega}_l}{2\pi})
\Gamma(\frac{1}{2}-\frac{\Delta_b}{2}-\frac{\bar{\omega}_l}{2\pi})}
-(l=0) \bigg] . \label{115}
\end{eqnarray}

One can show that
\begin{eqnarray}
\bigg[\frac{\Gamma(1-\frac{\Delta_f}{2}+\frac{\bar{\omega}_l}{2\pi})
\Gamma(1-\frac{\Delta_f}{2}-\frac{\bar{\omega}_l}{2\pi})}{(-1)^l
\Gamma(1-\Delta_f)} - (l=0) \bigg]
=\Big(\frac{1}{2}-\frac{\Delta_f}{2}\Big)
\frac{[\Gamma(\frac{1}{2}-\frac{\Delta_f}{2})]^2}{\Gamma(1-\Delta_f)}
\Bigg[ \frac{\prod_{k=1}^{l}(k+\frac{1}{2}-\frac{\Delta_f}{2})}
{\prod_{k=1}^{l}(k-\frac{1}{2}+\frac{\Delta_f}{2})} -1\Bigg] , \label{116} \\
\bigg[
\frac{(-1)^{l}\Gamma(-\Delta_b)}{\Gamma(\frac{1}{2}-\frac{\Delta_b}{2}+\frac{\bar{\omega}_l}{2\pi})
\Gamma(\frac{1}{2}-\frac{\Delta_b}{2}-\frac{\bar{\omega}_l}{2\pi})}
-(l=0) \bigg]  = -\Big(\frac{1}{2}-\frac{\Delta_f}{2}\Big)
\frac{\Gamma(-\Delta_b)}{[\Gamma(\frac{1}{2}+\frac{\Delta_f}{2})]^2}
\Bigg[ \frac{\prod_{k=1}^{l}(k+\frac{1}{2}-\frac{\Delta_f}{2})}
{\prod_{k=1}^{l}(k-\frac{1}{2}+\frac{\Delta_f}{2})} -1\Bigg] .
\label{117}
\end{eqnarray}
\end{widetext}

From Eqs. (\ref{115})-(\ref{117}) we get
\begin{eqnarray}
1&=& - \frac{V^2}{M} A_c A_f A_b (2 \pi)^2 \frac{\Gamma(-\Delta_b)
\Gamma(1-\Delta_f)}{\Big[\Gamma(\frac{1}{2}+\frac{\Delta_f}{2})
\Gamma(\frac{1}{2}-\frac{\Delta_f}{2})\Big]^2} \nonumber \\
&=& - \frac{V^2}{M} A_c A_f A_b \Gamma(-\Delta_b)
\Gamma(1-\Delta_f) [2 \sin(\pi \Delta_f /2)]^2 . \label{118} \nn
\end{eqnarray}

Equations (\ref{112}) and (\ref{118}) result in the following
equation
\begin{eqnarray}
M \Gamma(-\Delta_f) \Gamma(1-\Delta_b) = \Gamma(-\Delta_b)
\Gamma(1-\Delta_f) ,
\end{eqnarray}
or equivalently
\begin{eqnarray}
M \Delta_b = \Delta_f .
\end{eqnarray}
As a result, we obtain the solution
\begin{eqnarray}
\Delta_f &=& \frac{M}{M+1} , \\
\Delta_b &=& \frac{1}{M+1} ,
\end{eqnarray}
with the condition $M>1$, and $M = 2$ actually.

\subsubsection{Second solution: $\Delta_f=\Delta_b=1/2$}

In this case the scaling equation (\ref{93}) becomes
\begin{widetext}
\begin{eqnarray}
A_f^{-1} [\Phi_{f}(i\bar{\omega}_l)-\Phi_{f}(i\bar{\omega}_0)] &=&
-\frac{V^2}{M} A_c A_b [\Psi_{cb}(i\bar{\omega}_l)-
\Psi_{cb}(i\bar{\omega}_0)]  + J^2 A_f^3
[\Psi_{fff}(i\bar{\omega}_l)-\Psi_{fff}(i\bar{\omega}_0)] ,
\label{123}
\end{eqnarray}
\end{widetext}
where both the hybridization and RKKY interactions give rise to
the same order of magnitude for self-energy corrections. We can
use Eqs. (\ref{115})-(\ref{117}) with $\Delta_f=\Delta_b=1/2$,
where the RKKY term of $\Psi_{fff}(i\bar{\omega}_l)$ will give a
similar result.

\end{document}